\documentclass[12pt,a4]{article} 
\newcommand{\vs}{\vspace{-0.25cm}}
\usepackage{amsmath,graphicx}
\begin{document} 
\begin{center}
  {\Large{\bf Energy per particle of nuclear and neutron matter from subleading chiral
three-nucleon interactions}\footnote{This work 
has been supported in part by DFG and NSFC (CRC110).}  }  

\medskip

 N. Kaiser\\
\medskip
{\small Physik-Department T39, Technische Universit\"{a}t M\"{u}nchen,
   D-85747 Garching, Germany\\

\smallskip

{\it email: nkaiser@ph.tum.de}}
\end{center}
\medskip
\begin{abstract}
We derive from the subleading contributions to the chiral three-nucleon interaction [published in Phys.~Rev.~C77, 064004 (2008) and Phys.~Rev.~C84, 054001 (2011)] their first-order contributions to the energy per particle of isospin-symmetric nuclear matter and pure neutron matter in an analytical way. For the variety of short-range and long-range terms that constitute the subleading chiral 3N-force the pertinent closed 3-ring, 2-ring, and 1-ring diagrams are evaluated. While  3-ring diagrams vanish by a spin-trace and the results for 2-ring diagrams can be given in terms of elementary functions of the ratio Fermi-momentum over pion mass, one ends up in most cases for the closed 1-ring diagrams with one-parameter integrals. The same treatment is applied to the subsubleading chiral three-nucleon interactions as far as these have been constructed up to now. 
\end{abstract}

\section{Introduction}
Three-nucleon forces are an indispensable ingredient in accurate few-nucleon and
nuclear structure calculations. Nowadays, chiral effective field theory is the
appropriate tool to construct systematically the nuclear interactions in harmony with the symmetries of QCD. Three-nucleon forces appear first at next-to-next-to-leading order (N$^2$LO), where they consist of a zero-range contact-term ($\sim c_E$), a mid-range $1\pi$-exchange component ($\sim c_D$) and a long-range $2\pi$-exchange component ($\sim c_{1,3,4}$). The construction of the subleading chiral three-nucleon forces, built up by many pion-loop diagrams, has been performed for the long-range contributions in ref.\cite{3Nlong} and was completed with the short-range terms and relativistic $1/M$-corrections in ref.\,\cite{3Nshort}. Moreover, the extension of the chiral three-nucleon force to subsubleading order (N$^4$LO) has been acomplished  for the longest-range $2\pi$-exchange component in ref.\,\cite{twopi4} and for the intermediate-range contributions in ref.\,\cite{midrange4}. Very recently, the $2\pi$-exchange component of the 3N-force has also been analyzed in chiral effective field theory with $\Delta(1232)$-isobars as explicit degrees of freedom \cite{twopidelta} at order N$^3$LO. 

In order to implement these chiral 3N-forces into nuclear many-body calculations, a normal ordering to density-dependent NN-potentials has been performed by the Darmstadt group using a decomposition of the 3N-interaction with respect to  a $Jj$-coupled partial-wave momentum basis. The potentials obtained this way have been applied in second order many-body perturbation theory for calculations of the equation of state of isospin-asymmetric nuclear matter \cite{normalorder} and the energy per particle of pure neutron matter \cite{achim1}. In the same many-body framework the $nn$-pairing gaps in the $^1\!S_0$ and coupled $^3\!P_2$-$^3\!F_2$ channels \cite{achim2} have been computed, and the saturation properties of isospin-symmetric nuclear matter have been studied extensively\cite{achim3}. It is obvious that in this approach the treatment of the
chiral 3N-forces happens entirely in a numerical form through working with large data files for the 3N partial-wave matrix elements.

The purpose of the present work to provide results of an analytical calculation of the contributions to the energy per particle of isospin-symmetric nuclear matter, $\bar E(\rho)$, and pure neutron matter, $\bar E_n(\rho_n)$, as they are arise from the subleading chiral 3N-forces in their given (spin-isospin-momentum) operator form. We treat separately the variety of short-range terms and relativistic $1/M$-corrections derived in ref.~\cite{3Nshort} and the long-range terms comprising  $2\pi$-exchange, $1\pi2\pi$-exchange and ring topologies. One finds that a large fraction of the resulting contributions to the energies per particle can be written out in terms of arctangent or logarithmic functions (of the ratio Fermi-momentum over pion mass), while the rest is stated in the form of easily manageable integral representations. The present semi-analytical calculation is obviously restricted to first order many-body perturbation theory. However, the resulting expressions allow for a good check of the approximations induced by the normal-ordering procedure and the truncated partial wave sums in the purely numerical approaches. 
 
\section{Energy density from three-nucleon interactions}
\begin{figure}[ht]\centering
\includegraphics[width=0.3\textwidth]{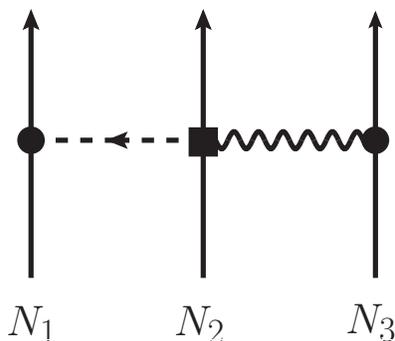}
\caption{Generic form of the 3N-interaction. The dashed line symbolizes 
pion-exchange and the wiggly line some other interaction.} \end{figure}
The generic form of the chiral 3N-interaction $V_\text{3N}$ is depicted in Fig.\,1. The dashed line symbolizes one-pion exchange and the wiggly line represents (exemplarily) a short-range interaction. However, any other interpretation of the wiggly line as a one-pion exchange or  a two-pion exchange is also possible. The essential feature of the symbolic diagram in Fig.\,1 is that one has a factorization of $V_\text{3N}$ in the three momentum transfers $\vec q_1, \vec q_2$ and $ \vec q_3$ at each nucleon line (satisfying the constraint $\vec q_1+\vec q_2+\vec q_3=0$). In the case of the ring-topology (see section 7) this factorization property is lost and an additional third (dashed) line connecting the left nucleon 1  and the right nucleon 3 is necessary for a complete illustration.  

\begin{figure}[ht]\centering
\includegraphics[width=0.8\textwidth]{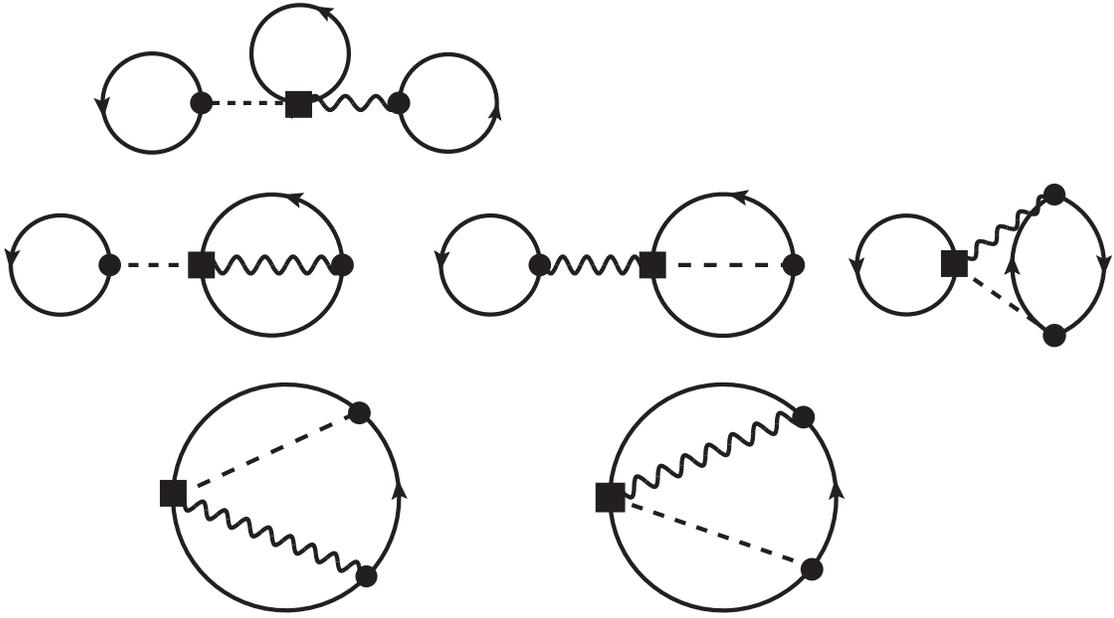}
\caption{Closed ring diagrams representing the energy density. There are: one 3-ring diagram, three 2-ring diagrams, and two 1-ring diagrams (all topologically distinct). Each nucleon-propagator carries a medium-insertion and thus loops are evaluated as Fermi sphere integrals:  $-(2\pi)^{-3}\!\int\!d^3p_j\, \theta(k_{f,n} -|\vec p_j|)$.}
\end{figure}

At first order in many-body perturbation the energy density of a filled nucleonic Fermi sea is represented by closed ring-diagrams that are obtained by concatenating the three nucleon lines of  $V_\text{3N}$. As shown in Fig.\,2, one gets one 3-ring diagram, three 2-ring diagrams, and two 1-ring diagrams that are topologically distinct. After taking spin- and isospin-traces over the closed rings, the three loops in momentum space are evaluated in the form of Fermi sphere integrals $-(2\pi)^{-3} \!\int\!d^3p_j\, \theta(k_f-|\vec p_j|)$. In the case of pure neutron matter only the spin-trace is present and one deals with three Fermi sphere integrals of the form $-(2\pi)^{-3}\!\int\!d^3p_j\,\theta(k_n-|\vec p_j|)$.  In the following sections and subsections we will first specify the detailed form of each (subleading chiral) 3N-interaction $V_\text{3N}$ and then give the corresponding contributions to the energy per particle of isospin-symmetric nuclear matter $\bar E(\rho)$ with density $\rho=2k_f^3/3\pi^2$, and pure neutron matter $\bar E_n(\rho_n)$ with density $ \rho_n=k_n^3/3\pi^2$. Let us note that the 3-ring diagrams vanish in almost all cases as a consequence of a spin-trace equal to zero. Whenever it is appropriate, the parts from the 2-ring diagrams and 1-ring diagrams will be specified separately.

\section{One-pion-exchange-contact topology}
For the $1\pi$-exchange-contact topology two nonzero contributions to $V_\text{3N}$ have been derived in section\,II of ref.\cite{3Nshort}, which take together the form:

\begin{equation} V_\text{3N}={g_A^4 C_Tm_\pi \over 32 \pi f_\pi^4} {\vec \sigma_1
\cdot \vec q_1\over m_\pi^2+ q_1^2} \big[2\vec\tau_1\!\cdot\!(\vec\tau_3-\vec\tau_2)
\vec \sigma_3\!\cdot\!\vec q_1 -\vec\tau_1\!\cdot\!(\vec\tau_2\!\times\!\vec
\tau_3)\, (\vec\sigma_2\!\times\! \vec\sigma_3)\!\cdot\! \vec q_1\big]\,.
\end{equation}
We remind that the parameter $C_T$ belongs to the leading order NN-contact potential $V_{\rm ct}= C_S+C_T\vec\sigma_1\!\cdot\!\vec\sigma_2$ and the factor $m_\pi/8\pi$ stems from  a pion-loop integral evaluated in dimensional regularization. The contributions from the 2-ring and 1-ring diagrams are equal and of opposite sign in the case of nuclear matter and for three neutrons the interaction in eq.(1) vanishes from the start ($\vec \tau_i\!\cdot \vec \tau_j \to 1, \vec\tau_1\!\cdot\!(\vec\tau_2\!\times\!\vec\tau_3)\to 0$):

\begin{equation} \bar E(\rho) = 0\,, \qquad \bar E_n(\rho_n) = 0\,. \end{equation}

\section{Two-pion-exchange-contact topology}
For the $2\pi$-exchange-contact topology ref.\cite{3Nshort} has derived two contributions to the 3N-interaction, where the structurally simpler one has the form:
\begin{equation}
V_\text{3N} = -{g_A^2 C_T \over 24 \pi f_\pi^4} \vec\tau_1\!\cdot\! \vec\tau_2
\,  \vec\sigma_2\!\cdot\! \vec\sigma_3 \big[m_\pi+(2m_\pi^2+q_1^2)A(q_1)\big]\,,
\end{equation}
with the pion-loop function $ A(q_1)=(1/2q_1)\arctan(q_1/ 2m_\pi)$.
A contribution to energy per particle of isospin-symmetric nuclear matter (of density $\rho=2k_f^3/3\pi^2$) arises only from the 1-ring diagrams:
\begin{equation}\bar E(\rho) ={g_A^2 C_T m_\pi^7 \over 560 \pi^5 f_\pi^4} \Big\{ 3u^2+{11u^4\over 2}-6u^6 -2u^5(7+3u^2)\arctan u -(3+7u^2) \ln(1+u^2)\Big\}\,, \end{equation}
with the dimensionless variable $u = k_f/m_\pi$. From the sum of the 2-ring and 1-ring diagrams one gets a contribution to energy per particle of pure neutron matter (of density $\rho_n=k_n^3 /3\pi^2$) which reads: 
\begin{equation}\bar E_n(\rho_n) ={g_A^2 C_T m_\pi^7 \over 1680 \pi^5 f_\pi^4} \Big\{ 3u^2+{11u^4\over 2}+{23u^6\over 2} -2u^5(7+3u^2)\arctan u -(3+7u^2) \ln(1+u^2)\Big\}\,, \end{equation}
with the ratio $u = k_n/m_\pi$. We make for the rest of the paper the agreement that in all formulas for $\bar E(\rho)$ the meaning of $u$ is $u = k_f/m_\pi$, whereas in all formulas for $\bar E_n(\rho_n)$ it means $u = k_n/m_\pi$.

The other 3N-interaction belonging to the $2\pi$-exchange-contact topology reads according to 
 ref.\,\cite{3Nshort}: 
\begin{equation} 
  V_\text{3N} = {g_A^4 C_T \over 48 \pi f_\pi^4} \bigg\{2 \vec\tau_1\cdot
\vec\tau_2\, \vec\sigma_2\cdot \vec\sigma_3\bigg[3m_\pi-{m_\pi^3\over 4
m_\pi^2+q_1^2}+2(2m_\pi^2+q_1^2)A(q_1)\bigg]  +9\big[\vec\sigma_1
\cdot\vec q_1\,\vec \sigma_2\cdot \vec q_1-q_1^2\, \vec\sigma_1\cdot\vec\sigma_2\big] A(q_1)\bigg\}\,,\end{equation}
where the first part is structurally equivalent to eq.(3). One obtains the following contributions to the energies per particle of nuclear and neutron matter:
\begin{equation}\bar E(\rho) ={g_A^4 C_T m_\pi^7 \over 1120 \pi^5 f_\pi^4} \Big\{ 3u^2-{17u^4\over 2}+36u^6 +4u^5(14+9u^2)\arctan u +(7u^2-3) \ln(1+u^2)\Big\}\,, \end{equation}
\begin{equation}\bar E_n(\rho_n) ={g_A^4 C_T m_\pi^7 \over 3360 \pi^5 f_\pi^4} \Big\{ 23u^2-{149u^4\over 2}+{43u^6\over 12} +4u^3(35+14u^2+6u^4)\arctan u -(23+77u^3) \ln(1+u^2)\Big\}\,. \end{equation}
\section{Leading relativistic corrections}
Next, one treats the relativistic $1/M$-corrections to the chiral 3N-interaction, which can be subdivided into diagrams of $1\pi$-exchange-contact topology (with parameter combination $g_A^2 C_{S,T}/f_\pi^2$) and  diagrams of $2\pi$-exchange topology (proportional to $g_A^2/
f_\pi^4$ or $g_A^4/ f_\pi^4$). Note that the corresponding expressions for $ V_\text{3N}$ depend on constants $\bar \beta_{8,9}$ which parametrize a unitary ambiguity of these 3N-potentials. In order to be consistent with the underlying NN-potential, one has to choose the values $\bar \beta_8 =1/4$ and $\bar \beta_9 =-1/4$ \cite{hebeler}. 
\subsection{$1\pi$-exchange-contact topology}
The $1/M$-correction to the $\pi$N-coupling combined with the 4N-contact vertex $(\sim C_{S,T}$) leads after setting $\bar \beta_9 =-1/4$ to a 3N-interaction of the form \cite{3Nshort}:
\begin{eqnarray}
 V_\text{3N}&=&-{g_A^2\over 16M f_\pi^2}{\vec\tau_1\!\cdot\! \vec\tau_2\over
m_\pi^2+q_1^2}\Big\{C_T\Big[i \vec\sigma_1\!\cdot\!(\vec p_1+\vec p_1\,\!\!')\,
(\vec  \sigma_2\!\times\!\vec \sigma_3)\!\cdot\! \vec q_1+3\vec\sigma_1\!\cdot\!
\vec q_1\,\vec\sigma_3\!\cdot\! \vec q_3\nonumber\\ &&\qquad \qquad \qquad
   \qquad +3i\vec \sigma_1
\!\cdot\!\vec q_1\,(\vec \sigma_2\!\times\!\vec\sigma_3)\!\cdot\!(\vec p_2+\vec
p_2\,\!\!') \Big] +3C_S \,\vec\sigma_1\!\cdot\! \vec q_1\, \vec\sigma_2\!\cdot\!
\vec q_3\Big\}\,,\end{eqnarray}
where $\vec p_j$ denotes ingoing momenta and $\vec p_j\,\!\!'$ outgoing momenta, such that $\vec q_j = \vec p_j\,\!\!'-\vec p_j$ are the momentum transfers. One finds the following contributions to the energies per particle of nuclear and neutron matter:
\begin{equation}\bar E(\rho) ={g_A^2(C_S-C_T) m_\pi^6 \over 512 \pi^4 M f_\pi^2} \bigg\{ 3u^2-18 u^4+8u^6 +24u^3\arctan 2u -\Big({3\over 4}+9u^2\Big) \ln(1+4u^2)\bigg\}\,, \end{equation}
\begin{equation}\bar E_n(\rho_n) ={g_A^2(C_S-3C_T) m_\pi^6 \over 512 \pi^4 M f_\pi^2} \bigg\{ u^2-6 u^4+{8u^6\over 3} +8u^3\arctan 2u -\Big({1\over 4}+3u^2\Big) \ln(1+4u^2)\bigg\}\,. \end{equation}
The retardation correction to the $1\pi$-exchange-contact diagram leads (after setting  $\bar \beta_8=1/4$) to a 3N-interaction of the form \cite{3Nshort}:
\begin{equation}
V_\text{3N}={g_A^2\over 16M f_\pi^2}{\vec\sigma_1\!\cdot\!\vec q_1\, \vec\tau_1\!\cdot\! 
\vec\tau_2\over (m_\pi^2+q_1^2)^2}\Big\{\vec q_1\!\cdot\!\vec q_3(C_S\,\vec\sigma_2\!\cdot\!
\vec q_1+C_T\,\vec\sigma_3\!\cdot\!\vec q_1)+i C_T(\vec\sigma_2\!\times\! 
\vec\sigma_3)\!\cdot\!\vec q_1\,(3\vec p_1+ 3\vec p_1\,\!'+\vec p_2+\vec p_2\,\!')\!\cdot 
\!\vec q_1\Big\}\,, \end{equation}
and it provides the following contributions to the energies per particle:
\begin{equation}\bar E(\rho) ={g_A^2(C_T-C_S) m_\pi^6 \over 128 \pi^4 M f_\pi^2} \bigg\{ u^2-3u^4+{2u^6\over 3} +5u^3\arctan 2u -{1\over 4}(1+9u^2) \ln(1+4u^2)\bigg\}\,, \end{equation}
\begin{equation}\bar E_n(\rho_n) ={g_A^2(3C_T-C_S) m_\pi^6 \over 384 \pi^4 M f_\pi^2} \bigg\{ u^2-3u^4+{2u^6\over 3} +5u^3\arctan 2u -{1\over 4}(1+9u^2) \ln(1+4u^2)\bigg\}\,. \end{equation}
It makes good sense that in pure neutron matter the contact potential $V_{\rm ct}= C_S+C_T\vec\sigma_1\!\cdot\!\vec\sigma_2$ shows up with its spin-singlet component $C_S-3C_T$, whereas in isospin-symmetric nuclear matter it enters through the average $C_S-C_T$ of its spin-singlet and spin-triplet parts.
\subsection{$2\pi$-exchange topology}
The $1/M$-correction to the isovector Weinberg-Tomozawa $\pi\pi$NN-vertex combined with two ordinary $\pi$N-couplings gives rise to a 3N-interaction of the form \cite{3Nshort}: 
\begin{equation} 2V_\text{3N}=-{g_A^2 \over 16M f_\pi^4}{\vec\sigma_1\!\cdot\! \vec q_1 \,\vec\sigma_3\!\cdot\! \vec q_3\over (m_\pi^2+q_1^2)(m_\pi^2+q_3^2)}\,\vec\tau_1\!\cdot\!(\vec\tau_2 
\!\times\! \vec\tau_3)\Big[\vec \sigma_2\!\cdot\!(\vec q_{1}\!\times\!\vec q_3)+{i \over 2}
(\vec p_ 2 +\vec p_2\,\!\!')\!\cdot\!(\vec q_3-\vec q_1)\Big]\,.
\end{equation}
which obviously vanishes for three neutrons $\vec\tau_1\!\cdot\!(\vec\tau_2 
\!\times\! \vec\tau_3)\to 0$. In those cases where the 3N-interaction $V_\text{3N}$ is symmetric under the exchange of nucleon 1 and nucleon 3, we multiply both sides with a factor 2. Note that when working with $2V_\text{3N}$ there is effectively only one 1-ring diagram and the right 2-ring diagram in Fig.\,2 carries a combinatoric factor $1/2$.  The nonvanishing contribution to the energy per particle of isospin-symmetric nuclear matter comes from the 1-ring diagram and its reads:  
\begin{equation} \bar E(\rho) = -{2g_A^2 m_\pi^6 \over (8\pi f_\pi)^4 M u^3} \int_0^u \!dx \Big\{ G(x)\Big[ (6u^2-3-7x^2)G(x) +8u^3x^2\Big] +[H(x)]^2 \Big\}\,,
\end{equation}
with auxiliary functions:
\begin{equation}G(x) = u(1+u^2+x^2) -{1\over 4x}\big[1+(u+x)^2\big] \big[1+(u-x)^2\big] \ln {1+(u+x)^2 \over 1+(u-x)^2}\,, \end{equation}
\begin{eqnarray} H(x) &\!\!\!=\!\!\!& ux(7u^2-x^2-13) +12x \big[\arctan(u+x)+ \arctan(u-x)\big] \nonumber \\ && + {1\over 4} \big[ (u^2-x^2)^2-11-10 u^2+14x^2\big] \ln {1+(u+x)^2\over 1+(u-x)^2}\,. \end{eqnarray} 
The retardation correction to the previous $2\pi$-exchange mechanism leads to a 3N-interaction of the form \cite{3Nshort}: 
\begin{equation}
2V_\text{3N}={ig_A^2\over 32M f_\pi^4}{\vec\sigma_1\!\cdot\! \vec q_1\,\vec\sigma_3\!\cdot\! \vec q_3 \over (m_\pi^2+q_1^2)(m_\pi^2+q_3^2)}\vec\tau_1\!\cdot\!(\vec\tau_2\!\times\!\vec\tau_3) \big[\vec q_3\!\cdot\!(\vec p_3+\vec p_3\!\,')-\vec q_1\!\cdot\!(\vec p_1+\vec p_1\!\,')\big]\,,
\end{equation}
which again vanishes for three neutrons. The corresponding contribution to energy per particle comes from the 1-ring diagram and it reads:
\begin{equation} \bar E(\rho) = {2g_A^2 m_\pi^6 \over (8\pi f_\pi)^4 M u^3} \int_0^u \!dx\, G(x)\Big[8u^3x^2+(6u^2-3-8x^2)G(x)-2x H(x)\Big]\,.  \end{equation}
The $1/M$-correction to the $\pi$N-coupling leads via the mechanism of two consecutive pion-exchanges to a 3N-interaction of the form \cite{3Nshort} (setting $\bar \beta_9=-1/4$):
\begin{eqnarray}
V_\text{3N}&=&{g_A^4 \over 64M f_\pi^4} {\vec\sigma_1\!\cdot\! \vec q_1 \over (m_\pi^2+q_1^2) 
(m_\pi^2+q_3^2)} \Big\{\vec \tau_1\!\cdot\! \vec\tau_3\big[3\vec\sigma_3\!\cdot\! \vec q_3\big(
               i\vec\sigma_2\!\cdot\!(\vec q_1\!\times\!(\vec p_2+\vec p_2\!\,'))+q_1^2\big)\\ &&+i\vec \sigma_3\!\cdot\!(\vec p_3+\vec p_3\!\,')\vec \sigma_2\!\cdot\!(\vec q_1\!\times\!\vec q_3)\big]-i\vec\tau_1\!\cdot\!(\vec\tau_2\!\times\! \vec\tau_3)\big[3\vec\sigma_3\!\cdot\! \vec q_3 \, \vec q_1\!\cdot\!(\vec p_2+\vec p_2\!\,')+\vec \sigma_3\!\cdot\!(\vec p_3+\vec p_3\!\,')\,\vec q_1 \!\cdot\!\vec q_3\big]\Big\}\,. \nonumber \end{eqnarray}
 It is advantageous to give the contributions from the 2-ring diagrams and  1-ring diagrams separately:
  \begin{equation}\bar E(\rho)^{2r} ={6g_A^4 m_\pi^6 \over ( 4\pi f_\pi)^4 M} \bigg\{ u^2-3u^4+{2u^6\over 3} +5u^3\arctan 2u -{1\over 4}(1+9u^2) \ln(1+4u^2)\bigg\}\,, \end{equation} 
  \begin{equation}\bar E_n(\rho_n)^{2r} ={g_A^4 m_\pi^6 \over (4\pi f_\pi)^4 M} \bigg\{ u^2-3u^4+{2u^6\over 3} +5u^3\arctan 2u -{1\over 4}(1+9u^2) \ln(1+4u^2)\bigg\}\,, \end{equation}   
           
\begin{eqnarray}\bar E(\rho)^{1r} &\!\!\!=\!\!\!&{3g_A^4 m_\pi^6 \over (8 \pi f_\pi)^4 Mu^3} \int_0^u \!\!dx \bigg\{2[Z(x)]^2 +{4u^3 \over 3}(5x^2-4-4u^2)G(x)\nonumber\\ &&+\bigg({(1+u^2)^2 \over x^2} -{5\over 2}-11x^2+2u^2\bigg) [G(x)]^2 +{64u^6 x^2\over 9} \bigg\}\,,  \end{eqnarray}

\begin{eqnarray}\bar E_n(\rho_n)^{1r} &\!\!\!=\!\!\!&{g_A^4 m_\pi^6 \over (8 \pi f_\pi)^4 Mu^3} \int_0^u \!\!dx \bigg\{[Z(x)]^2 +{4u^3 \over 3}(4+4u^2-5x^2)G(x)\nonumber\\ &&+\bigg({5\over 2}-2u^2-x^2 -{(1+u^2)^2 \over x^2}\bigg) [G(x)]^2 -{64u^6 x^2\over 9} \bigg\}\,,  \end{eqnarray}
 where $Z(x)$ is a new auxiliary function:
\begin{eqnarray} Z(x) &\!\!\!=\!\!\!& 2ux\Big({5u^2 \over 3}-5-x^2\Big) +8x \big[\arctan(u+x)+ \arctan(u-x)\big] \nonumber\\ && + \Big[{1\over 2} (u^2-x^2)^2-{3\over 2} -u^2+3x^2\Big] \ln {1+(u+x)^2\over 1+(u-x)^2}\,. \end{eqnarray} 
Finally, there is the retardation correction to the (consecutive) $2\pi$-exchange. It generates a 3N-interaction of the form \cite{3Nshort} (setting $\bar \beta_8=1/4$): 
\begin{eqnarray}
V_\text{3N}&=&{g_A^4 \over 64M f_\pi^4}{\vec\sigma_1\!\cdot\! \vec q_1\,\vec\sigma_3\!\cdot \! \vec q_3 \over (m_\pi^2+q_1^2)^2(m_\pi^2+q_3^2)}\Big\{-\vec q_1\!\cdot\!\vec q_3\big[\vec\tau_1  \!\cdot\! \vec\tau_3\,\vec q_1\!\cdot\!\vec q_3+\vec\tau_1\!\cdot\!(\vec\tau_2\!\times\!  \vec 
\tau_3)\, \vec\sigma_2\!\cdot\!(\vec q_1\!\times\! \vec q_3)\big] \nonumber \\ && +i\big[ \vec \tau_1 \!\cdot\! \vec\tau_3\, \vec\sigma_2\!\cdot\!(\vec q_1\!\times\! \vec q_3) -\vec\tau_1\!\cdot\!(\vec\tau_2\!\times\! \vec\tau_3)\, \vec q_1 \! \cdot\!\vec q_3 \big]\, \vec q_1\! 
\cdot\!(3\vec p_1 +3\vec p_1\!\,'+\vec p_2+\vec p_2\!\,')\Big\}\,.\end{eqnarray}
The corresponding contributions to the energies per particle as derived from the 2-ring diagrams read:
\begin{equation}\bar E(\rho)^{2r} ={g_A^4 m_\pi^6 \over (4\pi  f_\pi)^4 M} \bigg\{ 9u^4-5u^2-{4u^6\over 3} -{35u^3\over 2}\arctan 2u +\Big({5\over 4}+9u^2\Big) \ln(1+4u^2)\bigg\}\,, \end{equation}  
\begin{equation}\bar E_n(\rho_n)^{2r} ={g_A^4 m_\pi^6 \over 6(4\pi  f_\pi)^4 M} \bigg\{ 9u^4-5u^2-{4u^6\over 3} -{35u^3\over 2}\arctan 2u +\Big({5\over 4}+9u^2\Big) \ln(1+4u^2)\bigg\}\,, \end{equation} 
The more tedious evaluation of the 1-ring diagrams leads to one-parameter integrals of the form:
\begin{equation}\bar E(\rho)^{1r} ={3g_A^4 m_\pi^6 \over (8 \pi f_\pi)^4 Mu^3} \int_0^u \!\!dx \Big\{
8 G_s(x)Z_s(x)+4 G_t(x)Z_t(x) -{3\over 8}\big[ H_a(x)K_a(x)+  H_b(x)K_b(x)\big]\Big\}\,,\end{equation}
\begin{equation}\bar E_n(\rho_n)^{1r} ={g_A^4 m_\pi^6 \over (8 \pi f_\pi)^4 Mu^3} \int_0^u \!\!dx \Big\{
4 G_s(x)Z_s(x)-4 G_t(x)Z_t(x) +{3\over 8}\big[ H_a(x)K_a(x)+  H_b(x)K_b(x)\big]\Big\}\,,\end{equation}
with eight further auxiliary functions in order to obtain the integrands as nice sums of products: 
\begin{equation} G_s(x) =4 ux\Big({2u^2\over 3}-1\Big) +4x\big[\arctan(u+x)+ \arctan(u-x)\big] +(x^2-u^2-1) \ln {1+(u+x)^2\over 1+(u-x)^2}\,,\end{equation} 
\begin{equation} G_t(x) =ux\Big({4u^2\over 3}+{x^2 \over 2}\Big) -{u \over 2x}(1+u^2)^2 
+{1\over 8} \Big[{(1+u^2)^3 \over x^2} -x^4 +(1-3u^2)(1+u^2-x^2)\Big] \ln {1+(u+x)^2\over 1+(u-x)^2}\,,\end{equation} 
\begin{eqnarray} Z_s(x) &\!\!\!=\!\!\!& 2 ux\Big(7-{u^2\over 3}+x^2\Big) -10x\big[\arctan(u+x)+ \arctan(u-x)\big] \nonumber \\ && + {1\over 2} \big[ 3-(u^2-x^2)^2-8x^2\big]\ln {1+(u+x)^2\over 1+(u-x)^2}\,,\end{eqnarray} 
\begin{eqnarray} Z_t(x) &\!\!\!=\!\!\!&  5 ux\Big({1\over 2}+{u^2\over 3}\Big) +{u \over 2x}(1-u^2-2u^4)  
\nonumber \\ && +{1\over 8} \Big[{2u^6+3u^4-1 \over x^2}+(2u^2-5)x^2-6-6u^2 -4u^4 \Big] \ln {1+(u+x)^2 \over 1+(u-x)^2}\,,\end{eqnarray} 
\begin{equation} H_a(x) =u\Big[x^2+2-{2u^2\over 3}+{(1+u^2)^2 \over x^2}\Big] -{1\over 4 x^3}[1+(u+x)^2][1+(u-x)^2](1+u^2+x^2) \ln {1+(u+x)^2\over 1+(u-x)^2}\,,\end{equation} 
\begin{equation} H_b(x) =2u\Big[1+{5u^2\over 3}-{(1+u^2)^2 \over x^2}\Big] +{1\over 2 x^3}\big[(1+u^2)^2 +x^2-u^2x^2\big](1+u^2-x^2) \ln {1+(u+x)^2\over 1+(u-x)^2}\,,\end{equation} 
\begin{eqnarray} K_a(x) &\!\!\!=\!\!\!&  {u\over 4}\Big[{u^2\over 3}(2+11u^2)-3- {(1+u^2)^3 \over x^2}+\Big( {11u^2 \over 3}-3\Big) x^2 -x^4\Big]  \nonumber \\ && +{1\over 16x^3} \big[(1+u^2)^2+2x^2-2u^2x^2+x^4\big]^2  \ln {1+(u+x)^2 \over 1+(u-x)^2}\,,\end{eqnarray} 
\begin{eqnarray} K_b(x) &\!\!\!=\!\!\!&  {u\over 2}\Big[{(1+u^2)^3 \over x^2}-1 -{u^2\over 3}(14+11u^2)
+(7u^2-1)x^2+x^4 \Big]  \nonumber \\ && -{1\over 8x^3}[1+(u+x)^2][1+(u-x)^2](1+u^2-x^2)^2 \ln {1+(u+x)^2 \over 1+(u-x)^2}\,.\end{eqnarray} 
On can also devise a three-nucleon interaction induced by the Weinberg-Tomozawa $2\pi$-coupling at the central nucleon combined with pseudovector pion-couplings at the left and right nucleon. When evaluating the corresponding 1-ring diagram one obtains a contribution to the energy per particle of isospin-symmetric nuclear matter in the form of a relativistic $1/M$-correction:  
\begin{equation} \bar E(\rho) = -{18 g_A^2 m_\pi^6 \over (8\pi f_\pi)^4 M u^3} \int_0^u \!\!dx \, [G_s(x)]^2\,, \end{equation} 
with $G_s(x)$ written in eq.(32).

In the formulation of chiral effective field theory with explicit $\Delta(1232)$-isobar degrees of freedom also the first relativistic $1/M$-correction to the $2\pi$-exchange 3N-interaction has been derived in ref.~\cite{twopidelta}. It has the somewhat lengthy form (symmetric under $1\leftrightarrow 3$): 
\begin{eqnarray} 2V_\text{3N} &=&  {g_A^4 \over 64 Mf_\pi^4 \Delta^2} {\vec\sigma_1\!\cdot\! \vec q_1 \vec \sigma_3\!\cdot\! \vec q_3 \over (m_\pi^2+q_1^2) (m_\pi^2+q_3^2)}\Big\{\vec\tau_1 \!\cdot\! \vec\tau_3\Big[-8(\vec q_1\!\cdot\!\vec q_3)^2 \nonumber \\ && + i \vec\sigma_2\!\cdot\!(\vec q_1\!\times\!\vec q_3)\big[(\vec p_1+\vec p_1\,\!\!')\!\cdot\!(2 \vec q_1-\vec q_3) +(\vec p_3+\vec p_3\,\!\!')\!\cdot\!( \vec q_1-2\vec q_3)\big]\Big]  \nonumber \\ && + i\vec\tau_1\!\cdot\!(\vec\tau_2 \!\times\! \vec\tau_3)\vec q_1\!\cdot\!\vec q_3\Big[ (\vec p_1+\vec p_1\,\!\!')\!\cdot\!(\vec q_3-2\vec q_1)+(\vec p_3+\vec p_3\,\!\!')\!\cdot\!(2 \vec q_3-\vec q_1)+2i \vec\sigma_2\!\cdot\!(\vec q_1\!\times\!\vec q_3)\Big]\Big\}\,,\end{eqnarray}
with $\Delta = 293$\,MeV the delta-nucleon mass splitting. The evaluation of the (nonvanishing) right 2-ring diagram in Fig.\,2 with this expression for $V_\text{3N}$ gives the following contributions to the energies per particle:
\begin{equation}\bar E(\rho)^{2r} ={g_A^4 m_\pi^8 \over 64\pi^4  f_\pi^4 M \Delta^2} \bigg\{{5u^2\over 2}- 9u^4+{8u^6\over 3}-{8 u^8\over 5} +14u^3\arctan 2u -\Big({5\over 8}+6u^2\Big) \ln(1+4u^2)\bigg\}\,, \end{equation} 
\begin{equation}\bar E_n(\rho_n)^{2r} ={g_A^4 m_\pi^8 \over 384\pi^4  f_\pi^4 M \Delta^2} \bigg\{{5u^2\over 2}- 9u^4+{8u^6\over 3}-{8 u^8\over 5} +14u^3\arctan 2u -\Big({5\over 8}+6u^2\Big) \ln(1+4u^2)\bigg\}\,. \end{equation} 
At the same time the contributions arising from the 1-ring diagram evaluated with $2V_\text{3N}$  in eq.(41) can expressed as one-parameter integrals by introducing a new auxiliary function:
\begin{eqnarray} \Xi(x) &\!\!\!=\!\!\!& ux\Big({7u^2 \over 3}-11-{8u^4\over 5}-3x^2+{16 u^2x^2 \over 3}\Big) +8x\big[\arctan(u+x)+ \arctan(u-x)\big] \nonumber \\ && + {1\over 4} \big[ 3(u^2-x^2)^2-5 -2u^2+14x^2\big]\ln {1+(u+x)^2\over 1+(u-x)^2}\nonumber \\ &\!\!\!=\!\!\!&{2\over 3}H(x)-{7x\over 3}G(x)+8u^3x\Big({2x^2\over 3}-{u^2\over 5}\Big)\,,\end{eqnarray} 
in the following ways:
\begin{eqnarray}\bar E(\rho)^{1r} &\!\!\!=\!\!\!& {3g_A^4 m_\pi^8 \over (8 \pi f_\pi)^4 M\Delta^2 u^3} \int_0^u \!\!dx \bigg\{[\Xi(x)]^2 +{[G(x)]^2\over 64} \bigg[ {517 \over 2} +233u^2+{261 u^4\over 2} \nonumber \\ &&+ {15\over 4x^4}(1+u^2)^4 -{(1+u^2)^2 \over x^2}(13+51u^2)-(709+123u^2)x^2+{159x^4\over 4}\bigg]  \nonumber \\ &&+{u^3 \over 16} G(x) \bigg[ {1825 x^4\over 3}  -17\Big({61\over 3}+{277 u^2 \over 15}\Big) x^2 -{5 \over x^2}(1+u^2)^3 +{67 \over 3}+{274 u^2 \over 3}+69 u^4\bigg] \nonumber \\ && +{u^6 \over 6}\bigg[ {5\over 2}(1+u^2)^2 -\Big({41\over 3}+35u^2+{384 u^4\over 25}\Big) x^2 +\Big({1055\over 6}+{512 u^2\over  5}\Big) x^4 -{512 x^6\over 3}\bigg]\bigg\}\,, \end{eqnarray}
\begin{eqnarray}\bar E_n(\rho_n)^{1r} &\!\!\!=\!\!\!& {g_A^4 m_\pi^8 \over (8 \pi f_\pi)^4 M\Delta^2 u^3} \int_0^u \!\!dx \bigg\{{[\Xi(x)]^2\over 2} +{[G(x)]^2\over 64} \bigg[ {329 \over 2} -35u^2+{9 u^4\over 2} \nonumber \\ &&+ {75\over 4 x^4}(1+u^2)^4 -{(1+u^2)^2 \over x^2}(41+39u^2)+11(3u^2-19)x^2-{69x^4\over 4}\bigg]  \nonumber \\ &&+{u^3 \over 16} G(x) \bigg[ {677 x^4\over 3}  -\Big({673\over 3}+{1177 u^2 \over 15}\Big) x^2 -{25 \over x^2}(1+u^2)^3 +{239 \over 3}+{410 u^2 \over 3}+57 u^4\bigg] \nonumber \\ && +{u^6 \over 6}\bigg[ {25\over 2}(1+u^2)^2 -\Big({157\over 3}+31u^2+{192 u^4\over 25}\Big) x^2 +\Big({763\over 6}+{256 u^2\over  5}\Big) x^4 -{256 x^6\over 3}\bigg]\bigg\}\,. 
\end{eqnarray}
Here, quite some effort has been involved in the decomposition of the integrands into $[\Xi(x)]^2, [G(x)]^2$  and $G(x)$ which subsume all arctangent and logarithmic functions. 
\section{Long-range terms}
The long-range contributions to the subleading (and subsubleading) chiral 3N-interaction fall into two categories: $2\pi$-exchange and $1\pi 2\pi$-exchange, which will treated in the next three subsections.   
\subsection{$2\pi$-exchange topology}
According to eq.(2.9) in ref.\,\cite{3Nlong} the $2\pi$-exchange 3N-interactions reads: 
\begin{eqnarray}
2V_\text{3N}&=&{g_A^4 \over 128\pi f_\pi^6} {\vec\sigma_1\!\cdot\! \vec q_1 \vec
\sigma_3\!\cdot\! \vec q_3 \over (m_\pi^2+q_1^2) (m_\pi^2+q_3^2)}\Big\{\vec\tau_1
\!\cdot\! \vec\tau_3\big[m_\pi(m_\pi^2+q_1^2+q_3^2+2q_2^2) \nonumber \\ && +
(2m_\pi^2+q_2^2)(3m_\pi^2+q_1^2+q_3^2+2q_2^2)A(q_2) \big]   \nonumber \\ && +
\vec\tau_1\!\cdot\!(\vec\tau_2\!\times\! \vec\tau_3)\, \vec\sigma_2\!\cdot\!(
\vec q_1\!\times\!\vec q_3)\big[m_\pi+ (4m_\pi^2+q_2^2) A(q_2) \big]\Big\} \,, \end{eqnarray}
where we have multipled by a factor $2$, due to the symmetry of this $V_\text{3N}$ under $1\leftrightarrow 3$, and also the relation  $q_2^2 = q_1^2+q_3^2+2\vec q_1\!\cdot \!\vec q_3$ has been used. After evaluating the non-vanishing (right) 2-ring diagram in Fig.\,2 (obtained by closing $N_2$), one gets the following contributions to the energies per particle of nuclear and neutron matter: 
\begin{equation}\bar E(\rho)^{2r} ={3g_A^4 m_\pi^7 \over (4\pi)^5  f_\pi^6} \bigg\{{3u^2\over 4}- {7u^4\over 2}+{4u^6\over 3}+5u^3\arctan 2u -\Big({3\over 16}+2u^2\Big) \ln(1+4u^2)\bigg\}\,, \end{equation} 
\begin{equation}\bar E_n(\rho_n)^{2r} ={g_A^4 m_\pi^7 \over (4\pi)^5  f_\pi^6} \bigg\{{3u^2\over 8}- {7u^4\over 4}+{2u^6\over 3}+{5u^3\over 2}\arctan 2u -\Big({3\over 32}+u^2\Big) \ln(1+4u^2)\bigg\}\,. \end{equation} 
In the case of the contributions from the  1-ring diagram it is advantageous to consider first those terms in eq.(47) that do not involve the arctangent function $A(q_2)=(1/2q_2)\arctan(q_2/2m_\pi)$. For these pieces the contributions to the energies per particle can still be reduced to one-parameter integrals:
\begin{equation}\bar E(\rho)^{1r} ={3g_A^4 m_\pi^7 \over 8(4\pi)^5  f_\pi^6u^3} \int_0^u\!\!dx \bigg\{2 [G_s(x)]^2 +[G_t(x)]^2+ 3G(x)\Big[ {5\over 4 }G(x)-4u^3 x^2\Big] \bigg\} \,, \end{equation} \begin{equation}\bar E_n(\rho_n)^{1r} ={g_A^4 m_\pi^7 \over 8(4\pi)^5  f_\pi^6u^3} \int_0^u\!\!dx \bigg\{ [G_s(x)]^2 +2[G_t(x)]^2+ 3G(x)\Big[ {5\over 4 }G(x)-4u^3 x^2\Big] \bigg\} \,. \end{equation} 
For the remaining terms in eq.(47) proportional to $A(q_2)$ the evaluation of the 1-ring diagram leads to the expressions (involving one or four numerical integrations):
\begin{eqnarray}
\bar E(\rho)^{1r} &\!\!\!=\!\!\!& {3g_A^4 m_\pi^7 \over (4\pi)^5  f_\pi^6u^3} \bigg\{\int_0^u\!\!dx\, \arctan x \Big[(1\!+\!2x^2)\Big( (12x^2\!-\!1)\Gamma_g(x) -{8 u^3 x\over 3}(u\!-\!x)^2(2u\!+\!x)\Big) \nonumber \\ &&-(4x^2\!+\!3) \Gamma_h(x) \Big] + {3\over 8} \int_0^u\!\!dx\!\int_0^u\!\!dy\!\int_{|x-y|}^{x+y}\!\!dz\, x y (4\!-\!4z^2\!+\!z^4\!+\!z^6)  \Psi(x,y,z)\arctan{z \over 2}\bigg\}\,,\end{eqnarray}
\begin{eqnarray}\bar E_n(\rho_n)^{1r} &\!\!\!=\!\!\!& {g_A^4 m_\pi^7 \over (4\pi)^5  f_\pi^6u^3} \bigg\{\int_0^u\!\!dx\, (1+2x^2)\arctan x \Big[ (12x^2\!-\!1)\Gamma_g(x) -\Gamma_h(x)-{8 u^3 x\over 3}(u\!-\!x)^2(2u\!+\!x)\Big] \nonumber \\ && + {3\over 8} \int_0^u\!\!dx\!\int_0^u\!\!dy\!\int_{|x-y|}^{x+y}\!\!dz\, x y (2\!+\!z^2)^2(1\!+\!2z^2)  \Psi(x,y,z)\arctan{z \over 2}\bigg\}\,,\end{eqnarray}
where the functions $\Gamma_g(x)$, $\Gamma_h(x)$, and $\Psi(x,y,z)=\int_0^u\!d\xi\dots$ are defined at the end of the next subsection.
\subsection{Generic form of $2\pi$-exchange}
According to ref.~\cite{3Nfourth} the $2\pi$-exchange 3N-interaction can be written in the following general form, modulo terms of shorter range:
\begin{equation}2V_\text{3N}={g_A^2 \over 4f_\pi^4} {\vec\sigma_1\!\cdot\! \vec
q_1 \vec \sigma_3\!\cdot\! \vec q_3 \over (m_\pi^2+q_1^2) (m_\pi^2+q_3^2)}\big[
\vec\tau_1 \!\cdot\! \vec\tau_3\,\tilde g_+(q_2) +\vec\tau_1\!\cdot\!(\vec\tau_3
\!\times\! \vec\tau_2)\, \vec\sigma_2\!\cdot\!(\vec q_1\!\times\!\vec q_3)\,
\tilde h_-(q_2)\big]\,. \end{equation} 
Here, the two structure functions $\tilde g_+(q_2)$ and $\tilde h_-(q_2)$ are $f_\pi^2$ times the isoscalar non-spinflip and isovector spinflip $\pi N$-scattering amplitude at zero pion-energy $\omega=0$ and squared momentum-transfer $t= - q_2^2$. The corresponding expressions from chiral perturbation theory up to N$^4$LO can be found in eqs.(59,60) of ref.~\cite{kaisersingh}.
The evaluation of the non-vanishing 2-ring diagram (closing $N_2$) gives the contributions: 
\begin{equation}\bar E(\rho)^{2r} ={3g_A^2 m_\pi^4 \over 64\pi^4  f_\pi^4}\tilde g_+(0) \Big\{2u^4- u^2- 4u^3\arctan 2u +\Big({1\over 4}+2u^2\Big) \ln(1+4u^2)\Big\}\,, \end{equation} 
\begin{equation}\bar E_n(\rho_n)^{2r} ={g_A^2 m_\pi^4 \over 64\pi^4  f_\pi^4}\tilde g_+(0) \Big\{u^4- {u^2\over 2}- 2u^3\arctan 2u +\Big({1\over 8}+u^2\Big) \ln(1+4u^2)\Big\}\,. \end{equation} 
The evaluation of the 1-ring diagram proceeds in a way similar to eqs.(52,53) such that $\tilde g_+(q_2)$ and $\tilde h_-(q_2)$ remain under the integral together with certain weighting functions. For the contribution to the energy per particle of isospin-symmetric nuclear matter one gets the result: 
\begin{eqnarray}\bar E_(\rho)^{1r}&\!\!=\!\!&{3g_A^2 m_\pi^4 \over (2\pi f_\pi)^4 u^3} \bigg\{\int_0^u \!\!dx\, x\Big[\tilde g_+(2m_\pi x) \Gamma_g(x)+m_\pi^2 \tilde h_-(2m_\pi x)\Gamma_h(x)\Big] \nonumber\\ && + {3\over 8} \int_0^u\!\!dx\!\int_0^u\!\!dy\!\int_{|x-y|}^{x+y}\!\!dz\, x y z  \Big[(2+z^2) \tilde g_+(m_\pi z)+ m_\pi^2z^2(4+z^2)\tilde h_-(m_\pi z)\Big] \Psi(x,y,z)\bigg\}\,, \end{eqnarray}
whereas in the case of pure neutron matter the result is simpler:
\begin{equation}\bar E_n(\rho_n)^{1r}={g_A^2 m_\pi^4 \over (2\pi f_\pi)^4  u^3} \bigg\{\int_0^u \!\!dx\, x \tilde g_+(2m_\pi x) \Gamma_g(x) + {3\over 8} \int_0^u\!\!dx\!\!\int_0^u\!dy\!\int_{|x-y|}^{x+y}\!\!dz\, x y z (2+z^2) \, \tilde g_+(m_\pi z)\Psi(x,y,z)\bigg\}, \end{equation}
because of the absence of the $\pi N$-amplitude $\tilde h_-(q_2)$. The decomposition into $\Gamma_g(x), \Gamma_h(x)$ and $\Psi(x,y,z)$ follows from a partial fraction decomposition of the spin- and isospin-traced $V_\text{3N}$ with respect to pion-propagators. The pertinent auxiliary function that were encountered in the reduction of threefold Fermi-sphere integrals are: 
\begin{eqnarray}\Gamma_g(x) &\!\!\!=\!\!\!&{2ux \over 5}(u-x)(3u x+2x^2+2-13u^2)+\Big(x^2+4u^3x -{1\over 40}-{u^2\over 2}\Big) \arctan 2u \nonumber\\ && +\Big({1\over 40}+{u^2\over 2}-6u^2x^2-x^2+2x^4\Big)  
\arctan 2x \nonumber\\ && +\Big({1\over 40}+{u^2\over 2}-6u^2x^2-x^2+2x^4+4u^3 x\Big)\arctan(2u- 2x)\nonumber\\ && + \Big[x^2u\Big({3\over 2}+2u^2\Big)-{x\over 8}(1+12u^2)-{u^3 \over 2}-{2u^5\over 5} \Big] \ln(1+4u^2) \nonumber\\ && +\Big[{2x^5 \over 5}-x^3(1+2u^2)+{x\over 8}(1+12u^2)  \Big] \ln(1+4x^2) \nonumber\\ && +\Big[x^3(1+2u^2)-{2x^5 \over 5}-{x\over 8}(1+12u^2) -2u^3x^2+{u^3\over 2}+{2u^5\over 5} \Big] \ln[1+4(u-x)^2] \,, \end{eqnarray}

\begin{eqnarray}\Gamma_h(x) &\!\!\!=\!\!\!&{ux \over 35}(u-x)\big[2-22u^2-24u^4-9u x(3+4u^2)+x^2(41-940u^2)+216 u x^3+144 x^4\big]\nonumber\\ && +\Big[ {1\over 140} +{u^2\over 5}+x^2\Big({3\over 10}+2u^2\Big) +16 u^3 x^3\Big] \arctan 2u \nonumber\\ && +\Big[8x^6-24u^2x^4- x^2\Big({3\over 10}+2u^2\Big)-{u^2\over 5}-{1\over 140} \Big] \arctan 2x \nonumber\\ && +\Big[8x^6-24u^2x^4+16u^3x^3- x^2\Big({3\over 10}+2u^2\Big)-{u^2\over 5}-{1\over 140} \Big]\arctan(2u- 2x)\nonumber\\ && + \bigg[u^3\Big({1\over 4}+{2u^2\over 5}+{12u^4\over 35}\Big) -{16u^5 x^2\over 5}-x^3\Big({1\over 2}+6u^2\Big) +{3\over 4u}(1+4u^2)^2x^4\bigg] \ln(1+4u^2) \nonumber\\ && +x^3\Big[ 2u^2-{1\over 2}-{4x^2\over 5}(3+14u^2) +{72x^4\over 35}\Big]  \ln(1+4x^2)  +\bigg[ {16u^5 x^2\over 5} +x^3\Big( {1\over 2}-2u^2\Big) \nonumber\\ &&-12u^3x^4+ {4x^5\over 5}(3+14u^2) -{72x^7\over 35}-u^3\Big({1\over 4}+{2u^2\over 5}+{12u^4\over 35}\Big)\bigg] \ln[1+4(u-x)^2] \,, \end{eqnarray}
 \begin{equation}\Psi(x,y,z)= \int_0^u\!\!d\xi {\xi \over \sqrt{z^2P \!-\!(x^2\!-\!y^2)^2}} \ln {P +z^2\xi^2+2\xi \sqrt{z^2P\!-\!(x^2\!-\!y^2)^2}\over \sqrt{[(1\!+\!(x\!+\!\xi)^2][(1\!+\!(x\!-\!\xi)^2][(1\!+\!(y\!+\!\xi)^2][(1\!+\!(y\!-\!\xi)^2]}}\,, \end{equation} 
 with the polynomial $P = (1+x^2-\xi^2)(1+y^2-\xi^2)+(4+z^2)\xi^2$. Note that the function $\Psi(x,y,z)$ arises from a Fermi-sphere integral over the product of the two different pion-propagators (working with dimensionless momenta in units of the pion mass $m_\pi$). 
\subsection{$1\pi 2\pi$-exchange topology}
The $1\pi 2\pi$-exchange 3N-interaction arises from a large set of loop diagrams and according to ref.~\cite{midrange4} it can be written in the following general form:
\begin{eqnarray}
V_\text{3N}&=&{g_A^4 \over 256\pi f_\pi^6} { \vec\sigma_3\!\cdot\! \vec q_3 \over m_\pi^2+q_3^2}\Big\{ \vec\tau_1 \!\cdot\! \vec\tau_3\Big[\vec\sigma_2\!\cdot\! \vec q_1\,\vec q_1\!\cdot\! \vec q_3\, f_1(q_1) + \vec\sigma_2\!\cdot\! \vec q_1\,f_2(q_1)+ \vec\sigma_2\!\cdot\! \vec q_3\,f_3(q_1)\Big] \nonumber\\ &&   +\vec\tau_2\!\cdot\! \vec\tau_3\Big[\vec\sigma_1\!\cdot\! \vec q_1\,\vec q_1\!\cdot\! \vec q_3\, f_4(q_1) + \vec\sigma_1\!\cdot\! \vec q_3\,f_5(q_1)+\vec\sigma_2\!\cdot\! \vec q_1\,\vec q_1\!\cdot \!\vec q_3\, f_6(q_1)\nonumber\\ &&+\vec\sigma_2\!\cdot\! \vec q_1\, f_7(q_1)+ \vec\sigma_2\!\cdot\! \vec q_3\,\vec q_1\!\cdot \!\vec q_3\,f_8(q_1)+ \vec\sigma_2\!\cdot\! \vec q_3\,f_9(q_1)\Big] \nonumber\\ &&  +(\vec\tau_1\!\times \!\vec\tau_2)\!\cdot\! \vec\tau_3\, \Big[(\vec\sigma_1\!\times\!\vec\sigma_2)\!\cdot\!\vec q_1\Big(\vec q_1\!\cdot \!\vec q_3\, f_{10}(q_1)+f_{11}(q_1)\Big) + \vec \sigma_1\!\cdot\!(\vec q_1\!\times \!\vec q_3)\,\vec \sigma_2\!\cdot\!\vec q_1\, f_{12}(q_1)\Big] \Big\}\,,\end{eqnarray}
where the reduced functions $f_j(q_1)$ can be found in eqs.(2-11) of ref.~\cite{kaisersubsub}. When considering the earlier version of the $1\pi 2\pi$-exchange 3N-interaction of ref.\cite{3Nlong} the reduced functions should be taken from eqs.(12-16) of ref.~\cite{kaisersingh}, making some shifts of indices: $f_6\to f_7, f_7\to f_9, f_8\to f_{11}$.   

The evaluation of the non-vanishing 2-ring diagram (obtained by closing $N_1$) gives the following contributions to the energies per particle: 
\begin{equation}\bar E(\rho)^{2r} ={g_A^4 m_\pi^6 \over (4\pi)^5  f_\pi^6}f_9(0)\Big\{3u^4- {u^2\over 2 }- {4u^6 \over 3} -4u^3\arctan 2u +\Big({1\over 8}+{3u^2\over 2}\Big) \ln(1+4u^2)\Big\}\,, \end{equation} 
\begin{equation}\bar E_n(\rho_n)^{2r} ={g_A^4 m_\pi^6 \over \pi^5  (4f_\pi)^6}\big[f_3(0)+f_9(0) \big]\Big\{2u^4- {u^2\over 3 }- {8u^6 \over 9} -{8u^3\over 3}\arctan 2u +\Big({1\over 12}+u^2\Big) \ln(1+4u^2)\Big\}\,, \end{equation} 
where the relation $f_5(q)= -q^2 f_4(q)$ has been employed in the case of neutron matter. Furthermore, one obtains from both 1-ring diagrams the same amounts which read (after doubling)  for nuclear and neutron matter:
\begin{eqnarray}\bar E(\rho)^{1r} &\!\!\!=\!\!\!&{6g_A^4 m_\pi^6 \over(4\pi)^5
f_\pi^6 u^3} \int_0^u\!\! dx\, x\Big\{\big[f_2(q)\!+\! f_7(q)\!-\! 4f_{11}(q)\big] W_a(x)+ \big[ f_3(q)\!+\!f_5(q)\!+\!f_9(q)\!-\!2q^2f_{12}(q)\big] W_b(x) \nonumber\\ &&\qquad \qquad\qquad\qquad\,\, +\, m_\pi^2\big[f_1(q)\!+\!f_4(q)\!+\!  f_6(q)\!-\! 4f_{10}(q)\!+\! 2f_{12}(q)\big] W_c(x) \nonumber \\ && \qquad \qquad\qquad \qquad\,\,-\, m_\pi^2 f_8(q)\Big[ {8u^3 \over 3} x^3(u-x)^2(2u+x)+W_a(x)\Big]\Big\} \,, \end{eqnarray}
\begin{eqnarray}\bar E_n(\rho_n)^{1r} &\!\!\!=\!\!\!&{2g_A^4 m_\pi^6 \over(4\pi)^5
f_\pi^6 u^3} \int_0^u\!\! dx\, x\Big\{\big[f_2(q)\!+\! f_7(q)\big] W_a(x)+ \big[
f_3(q)\!+\!f_5(q)\!+\!f_9(q)\big] W_b(x) \nonumber\\ &&+\, m_\pi^2\big[f_1(q)\!+\!f_4(q)\!+\!  f_6(q)\big] W_c(x) - m_\pi^2 f_8(q)\Big[ {8u^3 \over 3} x^3(u-x)^2(2u+x)+W_a(x)\Big]\Big\} \,,\end{eqnarray}
where one has to set $q=2m_\pi x$. In the reduction of threefold Fermi-sphere integrals to one-parameter integrals the fact that all $f_j(q)$ are even functions of $q$ has been exploited.  The somewhat lengthy weighting functions $W_{a,b,c}(x)$, derived in the process of repeatedly changing the order of integrations, involve several arctangent and logarithmic functions and they read: 
\begin{eqnarray}W_a(x)&\!\!\!=\!\!\!&{ux \over 35}(u-x)\bigg[1-11u^2-12u^4-9u x\Big({3\over 2}+2u^2\Big)- \Big({71 \over 2}+106u^2\Big) x^2+24u x^3 +16x^4\bigg]\nonumber \\ && 
+\bigg({1\over 280}+{u^2\over 10}+{x^2\over 5} +2u^2x^2-2x^4\bigg)\big[\arctan2 u -\arctan2 x -\arctan(2 u-2x)\big] \nonumber \\ && +\bigg[{3x^4 \over 8u}+u^3\Big({1\over 8}+{u^2 \over 5}+{6u^4\over 35}+x^2-{4u^2x^2\over 5}+2x^4\Big) \bigg] \ln(1+4u^2)\nonumber \\ && + x^3\bigg( {8x^4 \over 35}-{8u^2x^2 \over 5}+{4x^2 \over 5}-2u^2-{1\over 2}\bigg) \ln(1+4x^2) +
\bigg[u^3 \Big({4u^2x^2\over 5}-x^2-{1\over 8}-{u^2\over 5}-{6u^4 \over 35}\Big) \nonumber \\ && +x^3\Big( {1\over 2}+2u^2-2u^3 x-{4x^2 \over 5}+{8u^2x^2\over 5}-{8x^4 \over 35}\Big) \bigg] \ln\big[1+4(u-x)^2\big] \,, \end{eqnarray}

\begin{eqnarray}W_b(x)&\!\!\!=\!\!\!& 2ux(u-x)\bigg[{1\over 5}(2-13u^2+3ux +2x^2) +{2u^2 \over 3}(2u^2-u x -x^2) \bigg]\nonumber \\ && + \bigg( 4u^3 x +x^2 -{u^2 \over 2}-{1\over 40}\bigg) 
\arctan2 u + \bigg(2x^4-6u^2 x^2-x^2 +{u^2 \over 2}+{1\over 40}\bigg)\arctan2 x \nonumber \\ &&
+\bigg(2x^4-6u^2 x^2-x^2 +4u^3 x +{u^2 \over 2}+{1\over 40}\bigg)\arctan(2 u-2x) \nonumber \\ && +\bigg[u x^2\Big({3\over 2}+2u^2\Big) -x\Big({1\over 8}+{3u^2 \over 2}\Big) -{u^3 \over 2}-{2u^5 \over 5} \bigg] \ln(1+4u^2)\nonumber \\ && +  x \Big[ {1\over 8} +{3u^2 \over 2} -x^2(1+2u^2)+{2x^4 \over 5}\Big] \ln(1+4x^2)\nonumber \\ &&  +\bigg[u^3 \Big({1\over 2}+{2u^2 \over 5}-2x^2\Big) -x\Big( {1\over 8}+{3u^2 \over 2}\Big) +x^3(1+2u^2)-{2x^5 \over 5} \bigg] \ln\big[1+4(u-x)^2\big]\,, 
\end{eqnarray}

\begin{eqnarray}W_c(x)&\!\!\!=\!\!\!& {x(u-x)\over 315}\bigg[2u-41 u^3-88u^5-80u^7- u^2x\Big( {89\over 2 }+112 u^2+120 u^4\Big) +u x^2\Big( {647\over 2}-2232 u^2 \nonumber \\ && +{4216 u^4 \over3}\Big)
  + x^3\Big( 105 \!+\!716 u^2\!-\!{1496 u^4 \over 3}\Big) +x^4\Big({105 \over u}\!+\!596 u \!+\!{1912 u^3 \over 3}\Big) -192 u^2 x^5 -128 u x^6\bigg] \nonumber \\ && +\bigg[ {1\over 70}\Big({1\over 18}+2u^2-x^2\Big) +{2x^2 \over 5}(4x^2-u^2) +{16u^3 x^3\over 3}\bigg]\big[\arctan2 u -\arctan(2u-2 x)\nonumber \\ &&- \arctan 2 x\big]+{8x^3 \over 3}(2u^3-3u^2 x+x^3) \arctan2 x+{8x^3 \over 3}(4u^3-3u^2 x+x^3) \arctan(2u-2x) \nonumber \\ && +\bigg[u^3\Big( {1\over 24}+{u^2\over 10}+ {6u^4\over 35}+{8u^6 \over 63}\Big)-u^3 x^2\Big( {1\over 2}+{4u^2 \over 5}+{24u^4 \over 35}\Big) -x^3\Big( {1\over 6}+2u^2\Big)  \nonumber \\ && +x^4\Big({8u^5\over 5}+2u^3+{3u \over 2}-{1\over 8u}\Big) +x^6\Big({1\over 12 u^3}+ {1\over 2u}-{8u^3 \over 3}\Big) \bigg]  \ln(1+4u^2)\nonumber \\ && +  x^3 \bigg[ {1\over 6} +2u^2 -{8x^2\over 5}(1+u^2)+{8x^4 \over 35}(1+8u^2)-{64 x^6 \over 315}\bigg] \ln(1+4x^2) \nonumber \\ &&  +\bigg[u^3 x^2\Big( {1\over 2}+{4u^2 \over 5}+{24u^4 \over 35} \Big) -u^3 \Big({1\over 24}+{u^2 \over 10}+{6u^4 \over 35}+{8u^6 \over 63}\Big) -x^3\Big( {1\over 6}+2u^2\Big) 
  \nonumber \\ && -{2u^3 \over 5}x^4(5+4u^2)+{8 \over 5}(1+u^2) x^5 +{8 u^3 x^6 \over 3}-{8 x^7\over 35}(1+8u^2)+{64 x^9 \over 315} \bigg] \ln\big[1+4(u-x)^2\big]\,.
\end{eqnarray}

\section{Ring topology} 
\subsection{Subleading order}
The three-nucleon ring interaction is generated by a circulating pion that gets absorbed and reemitted at each of the three nucleons. It possesses a rather complicated structure, because any factorization property in the three momentum transfers $\vec q_{1,2,3}$ is lost. We start with the basic expression for $ V_\text{3N}$ in the form of a three-dimensional loop-integral over pion-propagators and momentum-factors \cite{3Nlong}:
\begin{eqnarray}V_\text{3N}&\!\!\!\!=\!\!\!\!&{g_A^4\over 32f_\pi^6}\int\!{d^3l_2\over (2\pi)^3} {1\over (m_\pi^2+l_1^2)(m_\pi^2+l_2^2)(m_\pi^2+l_3^2)} \bigg\{2\vec \tau_1\!\cdot\!\vec \tau_2\Big[\vec l_1\!\cdot\!\vec l_2\, \vec l_2\!\cdot\!\vec l_3 -\vec\sigma_1\!\cdot\!(\vec l_2\!\times\!\vec l_3) \vec\sigma_3\!\cdot\!(\vec l_1\!\times\!\vec l_2)\Big]\nonumber\\ &&+\vec\tau_1\!\cdot\!(\vec \tau_2\!\times\!\vec \tau_3) \vec\sigma_1\!\cdot\!(\vec l_2\!\times\!\vec l_3)\, \vec l_1\!\cdot\!\vec l_2 +{g_A^2 \over m_\pi^2+l_2^2} \Big[-4\vec \tau_1\!\cdot\!\vec \tau_2\, \vec\sigma_2\!\cdot\!(\vec l_1\!\times\!\vec l_3) \vec\sigma_3\!\cdot\!(\vec l_1\!\times\!\vec l_2)\vec l_2\!\cdot\!\vec l_3 \\ && -2\vec \tau_1\!\cdot\!\vec \tau_3\, \vec l_1\!\cdot\!\vec l_2\, \vec l_1\!\cdot\!\vec l_3\,\vec l_2\!\cdot\!\vec l_3+ \vec\tau_1\!\cdot\!(\vec \tau_2\!\times\!\vec \tau_3) \vec\sigma_2\!\cdot\!(\vec l_1\!\times\!\vec l_3)\vec l_1\!\cdot\!\vec l_2\,\vec l_2\!\cdot\!\vec l_3 +3 \vec\sigma_1\!\cdot\!(\vec l_2\!\times\!\vec l_3)\vec\sigma_3\!\cdot\!(\vec l_1\!\times\!\vec l_2)\vec l_1\!\cdot\!\vec l_3\Big]\bigg\}\,,\nonumber \end{eqnarray}
where one has to set $\vec l_1= \vec l_2-\vec q_3$ and $\vec l_3= \vec l_2+\vec q_1$. 
The 3-ring diagram produces a nonvanishing contribution only in pure neutron matter, which reads: 
\begin{equation}  \bar E_n(\rho_n)^{3r} ={5 g_A^4 m_\pi k_n^6\over (4\pi)^5 f_\pi^6} \bigg( {7g_A^2 \over 9}-{2\over3}\bigg)\,,
\end{equation}
where the internal loop-integral has been evaluated in dimensional regularization, setting a linear divergence to zero: $\int_0^\infty\!\!dl\,1 = 0$. The 2-ring diagrams give for isospin-symmetric nuclear matter the contributions:
\begin{equation} \bar E(\rho)^{2r} ={3g_A^4 m_\pi^7 \over 1120 \pi^5 f_\pi^6}\bigg\{u^6-{31 u^4 \over 16}-{11u^2 \over 8}+u^5\Big({7\over 2}+u^2\Big) \arctan u+{1\over 8}(11+21u^2)\ln(1+u^2)\bigg\}\,,
\end{equation} 

\begin{equation} \bar E(\rho)^{2r} ={g_A^6 m_\pi^7 \over 640 \pi^5 f_\pi^6}\bigg\{{u^2\over 7}\Big( {617 u^2\over 4}-{88 u^4\over 3}-67\Big) -u^3\Big({95\over 2}+7u^2+{6u^4\over 7}\Big) \arctan u+\Big({67 \over 7}+{121u^2 \over 4}\Big)\ln(1+u^2)\bigg\}\,,
\end{equation} 
and for pure neutron matter the contributions:
\begin{equation} \bar E_n(\rho_n)^{2r} ={g_A^4 m_\pi^7 \over 1120 \pi^5 f_\pi^6}\bigg\{{u^2\over 8}\Big( {59 u^4 \over 3}-{599u^2 \over 12}-{1\over 2}\Big) +u^3\Big(u^4+{14 u^2 \over 3}+{35\over 4}\Big) \arctan u+{1\over 48}(3-119u^2)\ln(1+u^2)\bigg\}\,,
\end{equation} 

\begin{eqnarray} \bar E_n(\rho_n)^{2r} &\!\!\!=\!\!\!& {g_A^6 m_\pi^7 \over 4480 \pi^5 f_\pi^6}\bigg\{{u^2\over 12}\Big( {4457 u^2\over 4}-{769u^4 \over 3}-{943\over 2}\Big) +u^3\Big(u^4-{119 u^2 \over 6}-{1225\over 6}\Big) \arctan u \nonumber\\ && \qquad \qquad \quad +{1\over 24}(943+3143u^2)\ln(1+u^2)\bigg\}\,,\end{eqnarray}
where the parts proproportional to $g_A^4$ and $g_A^6$ have been written down separately. After taking the spin-trace, both 1-ring diagrams in Fig.\,2 contribute with equal amounts. From  the $g_A^4$-part of the 3N-ring interaction in eq.(70) one gets:
\begin{equation}  \bar E_n(\rho_n)^{1r} = {1\over 3} \bar E(\rho)^{1r}= {3 g_A^4 m_\pi^7\over 64f_\pi^6 u^3}\int\limits_{|\vec p_j|<u}\!\!\!{d^9 p \over (2\pi)^8} K_4\big(|\vec p_1\!-\!\vec p_3|,|\vec p_2\!-\!\vec p_1|,|\vec p_3\!-\!\vec p_2|\big)\,,\end{equation} 
with the integrand function $K_4(q_1,q_2,q_3)$ arising from a 3-dimensional loop-integral (with linear divergence dropped):
\begin{equation} 
K_4(q_1,q_2,q_3) = {2+q_2^2 \over \sqrt{\Sigma}} \arctan{\sqrt{\Sigma}\over 8+q_1^2+q_2^2+q_3^2} -{1\over q_1} \arctan{q_1\over 2} -{2 +q_2^2\over q_2} \arctan{q_2\over 2}-{1\over q_3} \arctan{q_3\over 2}- 2\,,\end{equation} 
where $\Sigma = (q_1q_2q_3)^2+(q_1+q_2+q_3)(q_1+q_2-q_3)(q_1+q_3-q_2)(q_2+q_3-q_1)$. The integral over three Fermi-spheres in eq.(76) can be parametrized by three radii $p_{1,2,3}\in[0,u]$, two directional cosines $x_{1,2}\in [-1,1]$ and one azimuthal angle $\varphi \in [0,\pi]$.  The third directional cosine is $x_3 = x_1 x_2 +\sqrt{(1-x_1^2)(1-x_2^2)}\cos\varphi$. The three trivial angular integrations   in  eq.(76) (together with a factor $2$ from the full $\varphi$-range)  provide a factor $16\pi^2$. When putting aside the first term in $K_4(q_1,q_2,q_3)$ that involves $\sqrt{\Sigma}$, one can actually solve all integrals and gets: 
\begin{equation} \bar E(\rho)^{1r}_{sol} = 3\bar E_n(\rho_n)^{1r}_{sol} ={g_A^4 m_\pi^7\over 1120\pi^5 f_\pi^6}\bigg\{ 11u^4-u^2-{3u^6\over 2} -u^3\Big( {35\over 2}+7u^2+{3u^4\over 2}\Big) \arctan u +(1+7u^2) \ln(1+u^2)\bigg\}\,.\end{equation} 
The 1-ring diagrams evaluated with the $g_A^6$-part in eq.(70) leads to similar results for the energies per particle:
\begin{equation} \bar E(\rho)^{1r}= {9 g_A^6 m_\pi^7\over 256f_\pi^6 u^3}\int\limits_{|\vec p_j|<u}\!\!\!{d^9 p \over (2\pi)^8} \Big\{{1\over 2}K_6\big(|\vec p_1\!-\!\vec p_3|,|\vec p_2\!-\!\vec p_1|,|\vec p_3\!-\!\vec p_2|\big)+\widetilde K_6\big(|\vec p_1\!-\!\vec p_3|,|\vec p_2\!-\!\vec p_1|,|\vec p_3\!-\!\vec p_2|\big)\Big\}\,,\end{equation} 
\begin{equation}  \bar E_n(\rho_n)^{1r} = {3 g_A^6 m_\pi^7\over 256f_\pi^6 u^3}\int\limits_{|\vec p_j|<u}\!\!\!{d^9 p \over (2\pi)^8} \Big\{{3\over 2}K_6\big(|\vec p_1\!-\!\vec p_3|,|\vec p_2\!-\!\vec p_1|,|\vec p_3\!-\!\vec p_2|\big)+\widetilde K_6\big(|\vec p_1\!-\!\vec p_3|,|\vec p_2\!-\!\vec p_1|,|\vec p_3\!-\!\vec p_2|\big)\Big\}\,,\end{equation} 
with an isoscalar kernel-function:
\begin{eqnarray}K_6(q_1,q_2,q_3)&\!\!\!=\!\!\!& \big(2+2q_2^2-q_1^2-q_3^2\big)\bigg[ {1\over q_1 } \arctan{q_1\over 2}+{1\over q_3} \arctan{q_3\over 2}\bigg]- {2+q_2^2 \over q_2} \arctan{q_2\over 2} \nonumber\\ &&+{2+q_2^2 \over 2\Sigma} \bigg\{q_1^4+q_2^4+q_3^4-2q_1^2q_2^2-2q_1^2q_3^2-2q_2^2q_3^2+{1 \over \sqrt{\Sigma}} \arctan{\sqrt{\Sigma}\over 8+q_1^2 +q_2^2+q_3^2} \nonumber\\ && \times \Big[ q_1^2q_3^2\Big((q_1^2+q_3^2)(8+ q_2^2) -16q_2^2 -3q_2^4\Big) -2\big(q_1^2+q_3^2-q_2^2\big)^3\Big]\bigg\}\,, \end{eqnarray}
and an isovector kernel-function:
\begin{eqnarray}\widetilde K_6(q_1,q_2,q_3)&\!\!\!=\!\!\!& \Big({6-q_2^2\over q_1}+2q_1\Big)  \arctan{q_1\over 2}+\Big({6-q_2^2\over q_3}+2q_3\Big)\arctan{q_3 \over 2}+ {2-q_2^2 \over q_2} \arctan{q_2\over 2} \nonumber\\ &&+{1\over \Sigma}\bigg\{(6+q_1^2+q_2^2+q_3^2)(8+q_1^2+q_2^2+q_3^2)\Big[{4\over 4+q_1^2} +{4\over 4+q_3^2}\Big] -{q_2^6 \over 2}+q_2^4(q_1^2+q_3^2-13) \nonumber\\ &&+q_2^2\Big[ 22(q_1^2+q_3^2)+13q_1^2q_3^2-{q_1^4+q_3^4\over 2} -28\Big] -16(6+q_1^2+q_3^2) -13(q_1^2-q_3^2)^2\bigg\} \nonumber\\ &&+{ 1\over 2\Sigma^{3/2}} \arctan {\sqrt{\Sigma}\over 8+q_1^2 +q_2^2+q_3^2} \Big\{q_2^6(q_1^2q_3^2-4)+q_2^4\Big[ 8+14q_1^2q_3^2+(q_1^2+q_3^2)(12+q_1^2q_3^2)\Big]\nonumber\\ &&-2q_2^2 \Big[(q_1^2+q_3^2)(12+q_1^2q_3^2)+6q_1^4+6q_3^4+ 4 q_1^2q_3^2\Big]+4(q_1^2-q_3^2)^2(4+q_1^2+q_3^2) \Big\}\,. \end{eqnarray}
Note that we have arranged for an integrand-function $\widetilde K_6(q_1,q_2,q_3)$ that is also symmetric under $q_1\leftrightarrow q_3$. For the terms in the first line of $K_6(q_1,q_2,q_3)$ and $\widetilde K_6(q_1,q_2,q_3)$, proportional to $\arctan(q_j/2)$, one can again solve all integrals and gets:
\begin{eqnarray}\bar E(\rho)^{1r}_{sol} &\!\!\!=\!\!\!&  {g_A^6 m_\pi^7 \over 640 \pi^5 f_\pi^6}\bigg\{{u^2\over 56}\Big(79-{5337u^2\over 10}+{409u^4 \over 15}\Big) +{u^3\over 4}\Big(65 +9u^2+{4u^4 \over 35}\Big) \arctan u  \nonumber\\ && \qquad \qquad \quad +\Big( u^4-{297 u^2\over 40}-{79\over 56}\Big)\ln(1+u^2)\bigg\}\,,\end{eqnarray}
\begin{eqnarray} \bar E_n(\rho_n)^{1r}_{sol} &\!\!\!=\!\!\!&{g_A^6 m_\pi^7 \over 1920 \pi^5 f_\pi^6}\bigg\{{u^2\over 56}\Big(145-{6983u^2\over 10}-{409u^4 \over 15}\Big) +{u^3\over 4}\Big(95 +23u^2-{4u^4 \over 35}\Big) \arctan u  \nonumber\\ && \qquad \qquad \quad -\Big( u^4+{503 u^2\over 40}+{145\over 56}\Big)\ln(1+u^2)\bigg\}\,.\end{eqnarray}

\subsection{Subsubleading order}
At subsubleading order (N$^4$LO) the 3N-ring interaction constructed in ref.~\cite{midrange4} involves the $\pi$N low-energy constants $c_{1,2,3,4}$ and one has three pieces distinguished by their dependence on $g_A^2$.
\subsubsection{Part proportional to $g_A^0$}
The 3N-ring interaction proportional to $g_A^0c_{1,2,3,4}$ is given by a euclidean 
loop-integral of the form: 
\begin{eqnarray}V_\text{3N}&=&-{1\over f_\pi^6}\int_0^\infty\!\! dl_0\!\int\!{ d^3l_2\over (2\pi)^4} {l_0^2\over (\bar m^2+l_1^2)(\bar m^2+l_2^2)(\bar m^2+l_3^2)} \nonumber\\ && \times \Big\{\vec \tau_2\!\cdot\!\vec \tau_3\big[2c_1 m_\pi^2+(c_2+c_3)l_0^2 +c_3 \,\vec l_2\!\cdot\!\vec l_3\big]+{c_4 \over 4}\vec\tau_1\!\cdot\!(\vec \tau_2\!\times\!\vec \tau_3) \,\vec \sigma_1\!\cdot\!(\vec l_3\!\times\!\vec l_2)\Big\}\,, \end{eqnarray}
with $\bar m = \sqrt{m_\pi^2+l_0^2}$ and one has to set $\vec l_1= \vec l_2-\vec q_3$ and $\vec l_3= \vec l_2+\vec q_1$. The evaluation of the closed 3-ring diagram in Fig.\,2 with this $V_\text{3N}$ gives a nonvanishing contribution ($\sim \rho_n^2$) only in the case  of pure neutron matter, that reads:
\begin{equation} \bar E_n(\rho_n)^{3r} ={m_\pi^2 k_n^6\over 6(2\pi f_\pi)^6}\bigg\{\Big( {4c_1\over 3}-c_2-2c_3\Big)\ln{m_\pi \over \lambda}+c_1 -{5\over 12}(c_2+2c_3)\bigg\}\,, \end{equation}
where the internal loop-integral has been regularized by a euclidean cutoff $\lambda$, and dropping the $\lambda^2$-divergence. The sum of the three 2-ring diagrams in Fig.\,2 evaluated with $V_\text{3N}$ in eq.(85) lead to the following contrubtions to the energies per particle of nuclear and neutron matter:
\begin{eqnarray} \bar E(\rho)^{2r} &\!\!\!=\!\!\!& {m_\pi^8 \over (2 \pi f_\pi)^6}\bigg\{u^6\Big[ c_3+ {c_2\over 2}-{2 c_1\over 3} + {u^2\over 30}(3c_2+4c_3)\Big]\ln{m_\pi \over \lambda}+ {u^2 \over 4}\Big(
{3 c_2\over 16} + {3 c_3\over 4}-c_1\Big) \nonumber\\ && +  {u^4 \over 4}\Big({47 c_1\over 3} -{397 c_2\over 80} - {289 c_3\over 20}\Big) +  {u^6 \over 3}\Big(2 c_1 - {53 c_2\over 40} - {31 c_3\over 10}\Big) -  {u^8 \over 300}\Big(31c_2+{169c_3\over 3}\Big) \nonumber\\ && +\bigg[{3 c_2\over 64} + {3 c_3\over 16}-{c_1\over 4}+{u^2\over 2} \Big({3 c_2\over 4} + {5 c_3\over 2}-3 c_1\Big) \bigg]\ln^2\!\big(u+\sqrt{1+u^2}\big)+\bigg[ {u\over 2}\Big( c_1 - {3 c_2\over 16} - {3 c_3\over 4}\Big) \nonumber\\ && + u^3\Big( {73 c_2\over 80}\!+\! {51 c_3\over 20}\!-\!{8 c_1\over 3}\Big)+ u^5\Big({9 c_2\over 20}\! +\! {14 c_3 \over 15}\!-\!{2 c_1\over 3}\Big)+ {u^7\over 30} (3 c_2\! + \!4 c_3)\bigg] \sqrt{1+u^2}\ln\!\big(u+\sqrt{1\!+\!u^2}\big)\bigg\}\,,  \nonumber\\ &&\end{eqnarray}
\begin{eqnarray} \bar E_n(\rho_n)^{2r} &\!\!\!=\!\!\!& {m_\pi^8 \over (2 \pi f_\pi)^6}\bigg\{u^6\bigg[ {c_2\over 4}+{c_3\over 2}-{ c_1\over 3} + u^2\Big({c_2\over 20}+{c_3\over 9}\Big)\bigg]\ln{m_\pi \over \lambda}+ {u^2 \over 8}\Big({3 c_2\over 16} + { c_3\over 4}-c_1\Big) \nonumber\\ && +  {u^4 \over 8}\Big({47 c_1\over 3} -{397 c_2\over 80} - {101 c_3\over 12}\Big) +  {u^6 \over 12}\Big(4 c_1 - {53 c_2\over 20} - 5 c_3\Big) -  {u^8 \over 120}\Big({67 c_2\over 10} + {133 c_3\over 9}\Big) \nonumber\\ && +{1\over 8}\bigg[{3 c_2\over 16} + {c_3\over 4}-c_1+u^2 \Big({3 c_2\over 2} + {7 c_3\over 3}-6 c_1\Big) \bigg]\ln^2\!\big(u+\sqrt{1+u^2}\big)+\bigg[ {u\over 16}\Big( 4c_1 - {3 c_2\over 4} - c_3\Big) \nonumber\\ && + u^3\Big( {73 c_2\over 160}\!+\! {19 c_3\over 24}\!-\!{4 c_1\over 3}\Big)+ u^5\Big({9 c_2\over 40}\! +\! {4 c_3 \over 9}\!-\!{ c_1\over 3}\Big)+ u^7\Big({c_2\over 20}\! + \!{c_3\over 9}\Big)\bigg] \sqrt{1+u^2}\ln\!\big(u+\sqrt{1+u^2}\big)\bigg\}\,.  \nonumber\\ &&\end{eqnarray}
These results can alternatively be obtained by employing $V_\text{med}^{(0)}$ in ref.~\cite{kaisersubsub,kaiserVmedneutron} as an effective two-body interaction (linear in density) that is then integrated of two Fermi-spheres. 
The contributions from both 1-ring diagrams (with equal amounts) read:
\begin{equation}  \bar E_n(\rho_n)^{1r} ={1\over 3}\bar E(\rho)^{1r}= -{3m_\pi^8 \over 8f_\pi^6 u^3}\int\limits_{|\vec p_j|<u}\!\!\!{d^9 p \over (2\pi)^9} \Big\{c_1 K_1(q_1,q_2,q_3)+{c_2 \over 8}K_2(q_1,q_2,q_3)+{c_3 \over 4}K_3(q_1,q_2,q_3)\Big\}\,,\end{equation} 
where one has to set $q_1=|\vec p_1\!-\!\vec p_3|$, $q_2=|\vec p_2\!-\!\vec p_1|$,  $q_3=|\vec p_3\!-\!\vec p_2|$ in the kernel-functions $K_{1,2,3}(q_1,q_2,q_3)$. In order to write out these new kernel-functions, one introduces the abbreviations $D=q_1^4+q_2^4+q_3^4-2q_1^2q_2^2-2q_1^2q_3^2 -2q_2^2q_3^2$ and $L(s) =(\sqrt{4+s^2}/s)\ln[(\sqrt{4+s^2}+s)/2]$ as well as the euclidean three-point function (in spectral representation):
\begin{equation}  J(q_1,q_2,q_3) = \int_2^\infty \!\!d\mu{\mu \over (\mu^2+q_2^2)\sqrt{g}} \ln{\mu(\mu^2+q_1^2+q_3^2)+\sqrt{(\mu^2-4)g}\over \mu(\mu^2+q_1^2+q_3^2 )-\sqrt{(\mu^2-4)g}} \,,\end{equation}
where $g = [\mu^2+(q_1+q_3)^2][\mu^2+(q_1-q_3)^2]$. Putting all the pieces together, one finds:
\begin{eqnarray}K_1(q_1,q_2,q_3) &\!\!\!=\!\!\!&  {1\over D}\Big[ q_1^2\big(q_2^2+q_3^2-q_1^2\big)L(q_1) +q_2^2\big(q_1^2+q_3^2-q_2^2\big)L(q_2) +q_3^2\big(q_1^2+q_2^2-q_3^2\big)L(q_3)\nonumber\\ &&+2\big(q_1^2q_2^2q_3^2-D\big) J(q_1,q_2,q_3)\Big]+{3\over 4}-\ln{m_\pi \over \lambda} \,,\end{eqnarray}
\begin{eqnarray}&& K_2(q_1,q_2,q_3) ={q_1^2q_2^2q_3^2 \over 2D}+ {1\over D^2}\bigg\{6(q_1^2q_2^2 q_3^2-D)^2 J(q_1,q_2,q_3) \nonumber\\ &&+ q_1^2\big(q_1^2-q_2^2-q_3^2 \big) L(q_1)\bigg[ {q_1^6\over 2}+q_1^4(5-q_2^2-q_3^2) +5 (q_2^2-q_3^2)^2+q_1^2\Big( {q_2^4+q_3^4\over 2}-4 q_2^2q_3^2-10(q_2^2+q_3^2)\Big)\bigg] \nonumber\\ && + q_2^2\big(q_2^2-q_1^2-q_3^2 \big) L(q_2)\bigg[ {q_2^6\over 2}+q_2^4(5-q_1^2-q_3^2) +5 (q_1^2-q_3^2)^2+q_2^2\Big( {q_1^4+q_3^4\over 2}-4 q_1^2q_3^2-10(q_1^2+q_3^2)\Big)\bigg] \nonumber\\ && + q_3^2\big(q_3^2-q_1^2-q_2^2 \big) L(q_3)\bigg[ {q_3^6\over 2}+q_3^4(5-q_1^2-q_2^2) +5 (q_1^2-q_2^2)^2+q_3^2\Big( {q_1^4+q_2^4\over 2}-4 q_1^2q_2^2-10(q_1^2+q_2^2)\Big)\bigg] \bigg\} \nonumber\\ &&+ \Big(6+{q_1^2+q_2^2+q_3^2\over 2}\Big) \ln{m_\pi \over \lambda} -4-{7\over 16}(q_1^2+q_2^2+q_3^2)\,,\end{eqnarray}

\begin{eqnarray} K_3(q_1,q_2,q_3)&\!\!\!=\!\!\!&  {1\over D} \bigg\{ 2(2+q_1^2)(D-q_1^2q_2^2 q_3^2) J(q_1,q_2, q_3)+L(q_1) q_1^2(2+q_1^2)(q_1^2-q_2^2-q_3^2) \nonumber\\ &&+ {L(q_2)\over 3} \bigg[q_2^6+
q_2^4 (q_1^2-2q_3^2+10)+4 (q_1^2-q_3^2)^2+q_2^2\Big(q_3^4-2q_1^4-5 q_1^2q_3^2-14(q_1^2+q_3^2)\Big)\bigg]\nonumber\\ &&+ {L(q_3)\over 3} \bigg[q_3^6 +q_3^4 (q_1^2-2q_2^2+10)+4 (q_1^2-q_2^2)^2+q_3^2\Big(q_2^4-2q_1^4-5 q_1^2q_2^2-14(q_1^2+q_2^2)\Big)\bigg]\bigg\}\nonumber\\ &&+ \Big(6+q_1^2+{q_2^2+q_3^2\over 3}\Big) \ln{m_\pi \over \lambda}-{19\over 6}-{11\over 36}(3q_1^2+ q_2^2+q_3^2)\,.\end{eqnarray}
One should note that the constant and polynomial pieces at the end of each formula are specific for our ultraviolet regularization by a euclidean cutoff $\lambda$ (and dropping the $\lambda^2$-divergence). As a  good check one verifies that contributions of the form $c_{1,2,3}m_\pi^2 k_n^6/(2\pi f_\pi)^6 \ln(m_\pi/\lambda)$ to $\bar E_n(\rho_n)$ vanish after summing the pieces from closed 3-ring, 2-ring, and 1-ring diagrams.  
\subsubsection{Part proportional to $g_A^2$}
The 3N-ring interaction proportional to $g_A^2c_{1,2,3,4}$ is given by a euclidean 
loop-integral of the form \cite{kaisersubsub}:
\begin{eqnarray}V_\text{3N}&=&-{g_A^2\over f_\pi^6}\int_0^\infty\!\! dl_0\!\int\!{ d^3l_2\over (2\pi)^4} {1\over (\bar m^2+l_1^2)(\bar m^2+l_2^2)(\bar m^2+l_3^2)} \Big\{\vec \tau_2\!\cdot\!\vec \tau_3\, \vec l_1\!\cdot\!(\vec l_2+\vec l_3)  \nonumber\\ && \times \big[2c_1 m_\pi^2+(c_2+c_3)l_0^2 +c_3 \,\vec l_2\!\cdot\!\vec l_3\big]+{c_4 \over 2}\Big[ \vec\tau_1\!\cdot\!(\vec \tau_2\!\times\!\vec \tau_3) \,\vec l_1\!\cdot\! \vec l_2 \,\vec \sigma_2\!\cdot\!(\vec l_1\!\times\!\vec l_3)+ \vec \tau_1\!\cdot\!(\vec\tau_2+\vec \tau_3)\nonumber\\ && \times \Big( \bar m^2(\vec \sigma_2\!\times\!\vec l_3)\!\cdot\!(\vec \sigma_3\!\times\!\vec l_2)  +\vec l_2\!\cdot\!\vec l_3\, \vec \sigma_2\!\cdot\!\vec l_1\, \vec \sigma_3\!\cdot\!\vec l_1 + \vec l_1\!\cdot\!\vec l_2\,\vec l_1\!\cdot\!\vec l_3 \, \vec \sigma_2\!\cdot\!\vec \sigma_3-2 \vec l_1\!\cdot\!\vec l_3\, \vec \sigma_2\!\cdot\!\vec l_2\, \vec \sigma_3\!\cdot\!\vec l_1  \Big) \Big]  \Big\}\,. \end{eqnarray}
Again, the closed 3-ring diagram provides only a contribution to the energy per particle of pure neutron matter:  
\begin{equation}  \bar E_n(\rho_n)^{3r} ={g_A^2 m_\pi^2 k_n^6\over 3(2\pi f_\pi)^6}\Big\{ (4c_1-c_2-6c_3)\ln{m_\pi \over \lambda}+3c_1 -{5\over 12}(c_2+6c_3)\Big\}\,. \end{equation}
The contributions from the three 2-ring diagrams are conveniently calculated  with the help of the in-medium potentials $V_\text{med}^{(0)}\sim g_A^2 c_{1,2,3,4}$ in ref.~\cite{kaisersubsub,kaiserVmedneutron} as:
\begin{eqnarray} \bar E(\rho)^{2r} &\!\!\!=\!\!\!& {g_A^2m_\pi^8 \over (2 \pi f_\pi)^6}\bigg\{u^6\Big[ c_2+6c_3- 4c_1 + {u^2\over 3}(c_2+4c_3)\Big]\ln{m_\pi \over \lambda}+ {3u^2 \over 8}\Big(4c_1-{c_2\over 4} - 3 c_3\Big) \nonumber\\ && +  {u^4 \over 8}\Big(20 c_1 -{7 c_2\over 4} - 17 c_3\Big) +  {u^6 \over 3}\Big(8 c_1 - {7c_2\over 4} - 13 c_3\Big) -  {u^8 \over 18}\Big(5c_2+29c_3\Big) \nonumber\\ && +\bigg[{3 \over 8}\Big( 4c_1-{c_2\over 4}-3c_3\Big)  + u^2 \Big(3c_1-{ c_2\over 4} - {5 c_3\over 2}\Big) \bigg]\ln^2\!\big(u+\sqrt{1+u^2}\big)+\bigg[ {3u\over 4}\Big( { c_2\over 4} + 3 c_3-4c_1\Big) \nonumber\\ && + u^3\Big( {3 c_2\over 8}+ { 7c_3\over 2}-4 c_1\Big)+ u^5\Big({5 c_2\over 6} + {16 c_3 \over 3}-4 c_1\Big)+ {u^7\over 3} ( c_2 + 4 c_3)\bigg] \sqrt{1+u^2}\ln\!\big(u+\sqrt{1+u^2}\big)\bigg\}\,,  \nonumber\\ &&\end{eqnarray}
\begin{eqnarray} \bar E_n(\rho_n)^{2r} &\!\!\!=\!\!\!& {g_A^2m_\pi^8 \over (2 \pi f_\pi)^6}\bigg\{u^6\bigg[ {c_2\over 2}+3c_3+{2c_4\over 3}- 2c_1 + {u^2\over 15}\Big({11c_2\over 6}+{32c_3\over 3}+2c_4\Big)\bigg]\ln{m_\pi \over \lambda}\nonumber\\ && + {u^2 \over 8}\Big({c_2\over 8} - {5c_3\over 3}- {5c_4\over 6}-2c_1\Big)+  {u^4 \over 4}\Big(33 c_1 -{829 c_2\over 240} - {901 c_3\over 90}+ {689 c_4\over 180}\Big)\nonumber\\ &&  +  u^6 \Big({16 c_1\over 9} - {47c_2\over 120} - {19 c_3\over 10}+{7 c_4\over 90}\Big) -  {u^8 \over 225}\Big({667c_2\over 24}+{526c_3\over 3}+{89c_4\over 4}\Big) \nonumber\\ && +{1\over 4}\bigg[ {c_2\over 16}-c_1-{5c_3\over 6}-{5c_4\over 12}  + u^2 \Big({ 5c_2\over 6}-10c_1 - { c_3\over 3}-2c_4\Big) \bigg]\ln^2\!\big(u+\sqrt{1+u^2}\big)\nonumber\\ && +\bigg[ {u\over 4}\Big(2c_1- { c_2\over 8} + {5 c_3\over 3} + {5 c_4\over 6}\Big)+ u^3\Big( {161 c_2\over 240}-6c_1+ { 107c_3\over 45}-{ 101c_4\over 180} \Big)\nonumber\\ && + {u^5\over 45}\Big({79 c_2\over 4} -90c_1+ 109 c_3 +7c_4\Big)+ {u^7\over 45} \Big( {11c_2\over 2} + 32 c_3+6c_4\Big)\bigg] \sqrt{1+u^2}\ln\!\big(u+\sqrt{1+u^2}\big)\bigg\}\,. \nonumber\\ &&\end{eqnarray} 
Finally, the 1-ring diagrams evaluated with $V_\text{3N}$ in eq.(94) lead to the result:
\begin{equation}  \bar E_n(\rho_n)^{1r} ={1\over 3}\bar E(\rho)^{1r}= -{3g_A^2m_\pi^8 \over 4f_\pi^6 u^3}\int\limits_{|\vec p_j|<u}\!\!\!{d^9 p \over (2\pi)^9} \Big\{c_1 K_1'({\bf q})+{c_2 \over 8}K_2'({\bf q}) +{c_3 \over 4}K_3'({\bf q})+{c_4 \over 4}K_4'({\bf q})\Big\}\,,\end{equation} 
where one has to set ${\bf q} = (q_1,q_2,q_3)$ with $q_1=|\vec p_1\!-\!\vec p_3|$, $q_2=|\vec p_2\!-\!\vec p_1|$,  $q_3=|\vec p_3\!-\!\vec p_2|$. The four kernel- functions (symmetric under $q_2\leftrightarrow q_3$) are given by the expressions: 
\begin{eqnarray}K_1'({\bf q}) &\!\!\!=\!\!\!&  {1\over D}\Big\{L(q_1)\big[ 3q_1^2(q_2^2+q_3^2)-q_1^4 -2 (q_2^2-q_3^2)^2\big] +L(q_2)\big[q_2^2(q_1^2+q_3^2)- (q_1^2-q_3^2)^2\big]  \nonumber\\ &&+
L(q_3)\big[q_3^2(q_1^2+q_2^2)- (q_1^2-q_2^2)^2\big]- J({\bf q})\big[2q_1^2q_2^2q_3^2+(2+q_2^2+q_3^2)D\big]\Big\} +{5\over 4}-3\ln{m_\pi \over \lambda} \,, \end{eqnarray}

\begin{eqnarray}K_2'({\bf q}) &\!\!\!=\!\!\!& {1\over D}\bigg\{2J({\bf q})\bigg[{3 q_1^2 q_2^2 q_3^2\over D} + 1 + q_2^2 + q_3^2\bigg](D-q_1^2q_2^2q_3^2)- {q_1^2 q_2^2 q_3^2\over 2 } \nonumber\\ &&+L(q_1) \bigg[{3 q_2^2 q_3^2\over D}\Big(q_1^2 (q_2^4 + 6 q_2^2 q_3^2 + q_3^4)-(q_2^2-q_3^2)^2(q_2^2 + q_3^2)\Big)+ {q_1^4 \over 6}  (10 + q_2^2 + q_3^2)  \nonumber\\ && +{q_1^6\over 6}- {q_1^2\over 3} \Big(q_2^4 + q_3^4 + q_2^2 q_3^2 + 13 (q_2^2 + q_3^2)\Big) + {8\over 3} (q_2^2-q_3^2)^2+3q_2^2q_3^2 (q_2^2+ q_3^2)\bigg] \nonumber\\ && +L(q_2)\bigg[ {3 q_1^2 q_3^2\over D}\Big(q_2^2 (q_1^4 + 6q_1^2 q_3^2 +q_3^4) - (q_1^2 - q_3^2)^2 (q_1^2 + q_3^2)\Big)+{q_2^4\over 6} (2-7q_1^2-q_3^2)  \nonumber\\ && + {5 q_2^6\over 6} + {q_2^2\over 3} \Big( q_1^4-2 q_3^4 + 4 q_1^2 q_3^2 - 5 (q_1^2 + q_3^2)\Big) +{4\over 3} (q_1^2-q_3^2)^2 + 3 q_1^2 q_3^2 (q_1^2 + q_3^2)\bigg]  \nonumber\\ &&+L(q_3)\bigg[ {3 q_1^2 q_2^2\over D}\Big(q_3^2 (q_1^4 + 6 q_1^2 q_2^2 +q_2^4) - (q_1^2 - q_2^2)^2 (q_1^2 + q_2^2)\Big)+{q_3^4\over 6} (2 - 7 q_1^2 - q_2^2) \nonumber\\ &&+ {5 q_3^6\over 6} +{q_3^2\over 3} \Big(q_1^4 - 2 q_2^4 + 4 q_1^2 q_2^2 - 5 (q_1^2 + q_2^2)\Big) + {4\over 3} (q_1^2 - q_2^2)^2 + 3 q_1^2 q_2^2 (q_1^2 + q_2^2)\bigg] \bigg\}\nonumber\\ && +\Big[6+{q_1^2\over 6}+{5\over 6}(q_2^2+q_3^2)\Big]\ln{m_\pi\over\lambda}-{7\over 3}-{1\over 144}(37q_1^2+95q_2^2+95q_3^2)\,,\end{eqnarray}

\begin{eqnarray}K_3'({\bf q}) &\!\!\!=\!\!\!&  {1\over D}\bigg\{(2+q_1^2)\Big[J({\bf q}) \Big(2q_1^2q_2^2 q_3^2 +(2+q_2^2+q_3^2)D\Big)+ L(q_1)\Big(q_1^4 -3q_1^2(q_2^2 + q_3^2) +2 (q_2^2-q_3^2)^2\Big)\Big]\nonumber\\ &&+L(q_2)\Big[q_2^2(2+q_1^2)(q_1^2+q_3^2-q_2^2)+\Big({7q_2^2\over 6}+{q_1^2+q_3^2\over 2}+{14\over 3}\Big)D\Big]  \nonumber\\ &&+L(q_3)\Big[q_3^2(2+q_1^2)(q_1^2+q_2^2-q_3^2)+\Big({7q_3^2\over 6}+{q_1^2+ q_2^2\over 2}+{14\over 3}\Big)D\Big]  \bigg\} \nonumber\\ && +\Big[18+2q_1^2+{5\over 3}(q_2^2+q_3^2)\Big]\ln{m_\pi \over \lambda} -{29\over 6}-{1\over 36}(48q_1^2+43q_2^2+43q_3^2)\,, \end{eqnarray}

\begin{eqnarray}K_4'({\bf q}) &\!\!\!=\!\!\!& {1\over D}\bigg\{J({\bf q}) \Big[(q_2^2+q_3^2-q_1^2)D+q_2^2 q_3^2\big((q_2^2-q_3^2)^2-q_1^2(q_2^2+q_3^2)\big)\Big]  \nonumber \\ && +q_2^2 L(q_2)\bigg[ {D\over 2} +q_3^2 (q_3^2-q_1^2-q_2^2)\bigg] +q_3^2 L(q_3)\bigg[ {D\over 2}+q_2^2(q_2^2-q_1^2-q_3^2)\bigg]\nonumber\\ && +L(q_1) \bigg[(16+7q_1^2){D\over 3} + {1\over 2 }(q_2^2+q_3^2) \big(q_1^4+(q_2^2-q_3^2)^2\big)-q_1^2(q_2^4+q_3^4) \bigg] \bigg\}\nonumber \\ && + \bigg(12+{7q_1^2\over 3}+q_2^2+q_3^2\bigg)\ln{m_\pi \over \lambda} -{7\over 3}-{17q_1^2\over 9}-{q_2^2+q_3^2\over 2}\,. \end{eqnarray}
Again, one verifies as good check that contributions of the form $g_A^2c_{1,2,3,4}m_\pi^2 k_n^6/(2\pi f_\pi)^6 \ln(m_\pi/\lambda)$ to $\bar E_n(\rho_n)$ vanish after summing the pieces from closed 3-ring, 2-ring, and 1-ring diagrams.
\subsubsection{Part proportional to $g_A^4$}
The 3N-ring interaction proportional to $g_A^4c_{1,2,3,4}$ is given by a euclidean 
loop-integral over three pion-propagators (one of them squared) times a long series of terms with different spin-, isospin-, and momentum dependence, which reads \cite{kaisersubsub}:
\begin{eqnarray}V_\text{3N}&=&{g_A^4 \over f_\pi^6}\int_0^\infty\!\! dl_0\!\int\!{ d^3l_2\over (2\pi)^4} {1\over (\bar m^2+l_1^2)(\bar m^2+l_2^2)^2(\bar m^2+l_3^2)} \nonumber\\ &&\times \Big\{\bar m^2\Big[ (\vec \sigma_1\!\times\!\vec l_3)\!\cdot \!(\vec \sigma_3\!\times\!\vec l_1)\big[ 2l_0^2(c_2+c_3)\vec \tau_1\!\cdot\!\vec \tau_3-6c_1 m_\pi^2 + \vec l_1\!\cdot\!\vec l_3 \big(c_4(\vec \tau_1\!+\!\vec \tau_3)\!\cdot\!\vec \tau_2- 3c_3\big)\big]\nonumber\\ && + 2 (\vec \sigma_1\!\times\!\vec l_2)\!\cdot \!(\vec \sigma_2\!\times\!\vec l_1)\big[ 2l_0^2(c_2+c_3) \vec \tau_1\!\cdot\!\vec \tau_2-6c_1 m_\pi^2 + \vec l_1\!\cdot\!\vec l_2 \big(c_4(\vec \tau_1\!+\!\vec \tau_2)\!\cdot\!\vec \tau_3- 3c_3\big)\big]\Big]\nonumber\\ && +{c_4\over 2} \vec l_1\!\cdot\!\vec l_2\big[ 2 \vec l_1\!\cdot\!\vec l_3 \,\vec \sigma_1\!\cdot\!(\vec l_3\!\times\!\vec l_2)-\vec l_2\!\cdot\!\vec l_3 \,\vec \sigma_2\!\cdot\!(\vec l_3\!\times\!\vec l_1)\big] \vec \tau_1\!\cdot\!(\vec \tau_2\!\times\!\vec \tau_3)+\vec l_1\!\cdot\!\vec l_2\,\vec l_1\!\cdot\!\vec l_3\, \vec l_2\!\cdot\!\vec l_3 \nonumber\\ && \times \big[ 2c_3 \vec \tau_1\!\cdot\!(2\vec \tau_2\!+\!\vec \tau_3) +2 \vec \sigma_2\!\cdot\!\vec \sigma_3\big( c_4 \vec \tau_1\!\cdot\!(\vec \tau_2\!+\!\vec \tau_3)-3c_3\big) +\vec \sigma_1\!\cdot\!\vec \sigma_3\big( c_4(\vec \tau_1\!+\!\vec \tau_3)\!\cdot\!\vec \tau_2 -3c_3\big)\big]  \nonumber\\ && +2 \vec l_1\!\cdot\!\vec l_2\,\vec l_1\!\cdot\!\vec l_3\big[ (c_2+c_3)l_0^2( 2 \vec \sigma_2\!\cdot\!\vec \sigma_3\,\vec \tau_2\!\cdot\!\vec \tau_3-3)-6c_1 m_\pi^2\vec \sigma_2\!\cdot\!\vec \sigma_3 +\vec \sigma_1\!\cdot\!\vec l_3\, \vec \sigma_2\!\cdot\!\vec l_2\big( 3c_3-c_4(\vec \tau_1\!+\!\vec \tau_2)\!\cdot\!\vec \tau_3\big) \big] \nonumber\\ && + \vec l_1\!\cdot\!\vec l_2\,\vec l_2\!\cdot\!\vec l_3\big[4c_1 m_\pi^2\vec \tau_1\!\cdot\!\vec \tau_3-3 l_0^2(c_2+ c_3)+ 2\vec \sigma_1\!\cdot\!\vec \sigma_3\big(l_0^2(c_2+ c_3)\vec \tau_1\!\cdot\!\vec \tau_3-3c_1 m_\pi^2\big)
+ 2\vec \sigma_1\!\cdot\!\vec l_1\,\vec \sigma_2\!\cdot\!\vec l_3 \nonumber\\ &&  \times \big( 3c_3-c_4(\vec \tau_1\!+\!\vec \tau_2)\!\cdot\!\vec \tau_3 \big) \big] + 2\vec l_1\!\cdot\!\vec l_3\,\vec l_2\!\cdot\!\vec l_3\big[4c_1 m_\pi^2\vec \tau_1\!\cdot\!\vec \tau_2 +\vec \sigma_1\!\cdot\!\vec l_1\,\vec \sigma_3\!\cdot\!\vec l_2\big(3c_3-c_4(\vec \tau_1\!+\!\vec \tau_3)\!\cdot\!\vec \tau_2 \big) \big] \nonumber\\ &&+(\vec l_1\!\cdot\!\vec l_3)^2 \vec \sigma_1\!\cdot\!\vec l_2\,\vec \sigma_3\!\cdot\!\vec l_2\big( c_4(\vec \tau_1\!+\!\vec \tau_3)\!\cdot\!\vec \tau_2 -3c_3\big)+ 2(\vec l_1\!\cdot\!\vec l_2)^2 \vec \sigma_1\!\cdot\!\vec l_3\,\vec \sigma_2\!\cdot\!\vec l_3\big( c_4(\vec \tau_1\!+\!\vec \tau_2)\!\cdot\!\vec \tau_3 -3c_3\big)\nonumber\\ &&+4\big( \vec l_2\!\cdot\!\vec l_3\, \vec \sigma_1\!\cdot\!\vec l_1\,\vec \sigma_2\!\cdot\!\vec l_3 +\vec l_1\!\cdot\!\vec l_3\, \vec \sigma_1\!\cdot\!\vec l_3\,\vec \sigma_2\!\cdot\!\vec l_2-  \vec l_1\!\cdot\!\vec l_2\, \vec \sigma_1\!\cdot\!\vec l_3\,\vec \sigma_2\!\cdot\!\vec l_3\big)\big( 3c_1m_\pi^2 - l_0^2 (c_2+c_3)  \vec \tau_1\!\cdot\!\vec \tau_2\big)\nonumber\\ &&+2\big( 2 \vec l_2\!\cdot\!\vec l_3\, \vec \sigma_1\!\cdot\!\vec l_1 -\vec l_1\!\cdot\!\vec l_3\, \vec \sigma_1\!\cdot\!\vec l_2\big)\vec \sigma_3\!\cdot\!\vec l_2 \big( 3c_1m_\pi^2 - l_0^2 (c_2+c_3) \vec \tau_1\!\cdot\!\vec \tau_3\big)\Big\}\,, \end{eqnarray}
with $\bar m = \sqrt{m_\pi^2+l_0^2}$ and one has to set $\vec l_1= \vec l_2-\vec q_3$ and $\vec l_3= \vec l_2+\vec q_1$. For the first time one gets  from the closed 3-ring diagram evaluated with $V_\text{3N}$ in eq.(103) contributions to both the energy per particle of isospin-symmetric nuclear matter and pure neutron matter: 
\begin{equation} \bar E(\rho)^{3r} ={5g_A^4m _\pi^2 k_f^6\over (2\pi f_\pi)^6}(c_2+c_3)\bigg[ - \ln{m_\pi \over \lambda} -{13\over 24}\bigg]\,, \end{equation}
\begin{equation} \bar E_n(\rho_n)^{3r} ={5g_A^4 m_\pi^2 k_n^6\over 12(2\pi f_\pi)^6}\bigg\{ (11c_3-3c_2-8c_1)\ln{m_\pi \over \lambda}+{143c_3\over 24}-{13c_2\over 8}-{22c_1\over 3}\bigg\}\,. \end{equation}
The contributions from the three 2-ring diagrams are again conveniently calculated  with the help of the in-medium potentials $V_\text{med}^{(0)}\sim g_A^4 c_{1,2,3,4} $ in ref.~\cite{kaisersubsub,kaiserVmedneutron} as:
\begin{eqnarray} \bar E(\rho)^{2r} &\!\!\!=\!\!\!& {g_A^4m_\pi^8 \over (2 \pi f_\pi)^6}\bigg\{-{u^6\over 4}\Big[8c_1+9 c_2+19c_3 + {u^2\over 15}(3c_2+133c_3)\Big]\ln{m_\pi \over \lambda}+ {u^2 \over 4}\Big(25c_1-{35c_2\over 32} - {789 c_3\over 32}\Big) \nonumber\\ && +  {u^4 \over 4}\Big({2471 c_2\over 480}- {119c_1\over 3} +{17947 c_3\over 160}\Big) +  u^6 \Big({c_1\over 3} - {269c_2\over 480} + {407 c_3\over 160}\Big) +  {u^8 \over 1200}\Big(7c_2+{5771c_3\over 3}\Big) \nonumber\\ && +{1\over 4}\bigg[25c_1-{35c_2\over 32}-{789c_3\over 32}  + u^2 \Big(54c_1-{15 c_2\over 4} - {301 c_3\over 4}\Big) \bigg]\ln^2\!\big(u+\sqrt{1+u^2}\big)\nonumber\\ && +\bigg[ {u\over 2}\Big( {35c_2\over 32}-25c_1+{789c_3\over 32}\Big)+ u^3\Big({8 c_1\over 3}- {299 c_2\over 480}- { 2463c_3\over 160}\Big)+ {u^5\over 3}\Big(2c_1+{13 c_2\over 40} - {557 c_3 \over 40}\Big)\nonumber\\ && - {u^7\over 60} ( 3c_2 + 133 c_3)\bigg] \sqrt{1+u^2}\ln\!\big(u+\sqrt{1+u^2}\big)\bigg\}\,, \end{eqnarray}

\begin{eqnarray} \bar E_n(\rho_n)^{2r} &\!\!\!=\!\!\!& {g_A^4m_\pi^8 \over (2 \pi f_\pi)^6}\bigg\{ u^6\Big[{7c_2 \over 8}-c_1-{3 c_3\over 8}-5c_4 + {u^2\over 360}(107c_2-313c_3-180c_4)\Big]\ln{m_\pi \over \lambda}\nonumber\\ && + {u^2 \over 12}\Big(13 c_1 - {37 c_2\over 64} - {3937 c_3\over 64} + {351 c_4 \over 16}\Big)  +  {u^4 \over 3}\Big({252541 c_3\over 3840}- {14879c_2\over 3840} -{77c_1\over 3}-{1109 c_4\over 64}\Big)\nonumber\\ &&  +  u^6 \Big({1097 c_3\over 960} - {231 c_2\over 320} -2 c_1 + {23 c_4\over 72}\Big) +  {u^8 \over 21600}\big(12911c_3-4957c_2+6023c_4\big) \nonumber\\ && +{1\over 4}\bigg[{13c_1\over 3}-{37c_2\over 192}-{3937c_3\over 192}+{117c_4\over 16}  + u^2 \Big(17c_1\!+\!{11 c_2\over 24} \!-\! {1507 c_3\over 24}\!+\!22c_4\Big) \bigg]\ln^2\!\big(u+\sqrt{1+u^2}\big)\nonumber\\ && +\bigg[ {u\over 32}\Big({37 c_2\over 12} -{208 c_1\over 3} + {3937 c_3\over 12} - 117 c_4 \Big)+ {u^3\over 48} \Big({776 c_1\over 3} + {3251 c_2\over 60} - {32689 c_3\over 60} + 101c_4\Big)+{u^5\over 12}\nonumber\\ && \times\Big({44 c_1\over 3} \!+\! {241 c_2\over 20}\! -\! {319 c_3\over 20} \!-\! 25 c_4\Big)\ + {u^7\over 360} ( 107 c_2 \!- \!313 c_3 \!- \!180 c_4)\bigg] \sqrt{1+u^2}\ln\!\big(u+\sqrt{1+u^2}\big)\bigg\}\,.\nonumber\\ &&  \end{eqnarray}
The contributions from both 1-ring diagrams (with equal share) can be written in the form:
\begin{equation}  \bar E(\rho)^{1r}= {9g_A^4m_\pi^8 \over 8f_\pi^6 u^3}\int\limits_{|\vec p_j|<u}\!\!\!{d^9 p \over (2\pi)^9} \Big\{c_1\big({\cal  K}_1+\widetilde{\cal K}_1\big)+{c_2+c_3 \over 4}\big({\cal  K}_2+\widetilde{\cal  K}_2\big) +{c_3 \over 4}\big({\cal  K}_3+\widetilde{\cal  K}_3\big)-{c_4 \over 2}{\cal  K}_3\Big\}\,,\end{equation} 
\begin{equation}  \bar E_n(\rho_n)^{1r}= {3g_A^4m_\pi^8 \over 8f_\pi^6 u^3}\int\limits_{|\vec p_j|<u}\!\!\!{d^9 p \over (2\pi)^9} \Big\{c_1\big(3{\cal  K}_1+\widetilde{\cal  K}_1\big)+{c_2 +c_3\over 4}\big(3{\cal  K}_2+\widetilde{\cal  K}_2\big) +{c_3 \over 4}\big(3{\cal  K}_3+\widetilde{\cal  K}_3\big)-{c_4 \over 2}{\cal  K}_3\Big\}\,,\end{equation} 
where for simplicity the argument ${\bf q}=(q_1,q_2,q_3)$ on the three isoscalar kernel-functions ${\cal  K}_{1,2,3}({\bf q})$ and  three isovector kernel-functions $\widetilde{\cal  K}_{1,2,3}({\bf q})$ has been dropped. These are constructed such that they are symmetric under the exchange  of variables $q_1\leftrightarrow q_3$ . Introducing the abbreviation $\Sigma = (q_1 q_2 q_3)^2-D$ for an additionally occurring denominator,  one finds for the pair associated with $c_1$  the following  expressions:
\begin{eqnarray}{\cal  K}_1({\bf q})&\!\!\!=\!\!\!& {q_3^2-q_2^2\over 2q_1^2}+{q_1^2-q_2^2\over 2q_3^2}+J({\bf q})\bigg[4+3q_2^2-q_1^2-q_3^2+{2\over D}(q_1^2+q_3^2)\Big((q_1^2-q_3^2)^2-q_2^2(q_1^2+q_3^2)\Big)\bigg] \nonumber\\ && +L(q_2)\bigg\{{4q_2^2\over D}( q_1^2+q_3^2)-1-q_1^2-q_3^2+{1\over \Sigma}\Big[q_2^2\big(6q_1^2+6q_3^2+2q_1^4+2q_3^4+7 q_1^2q_3^2+q_1^4q_3^2+q_1^2q_3^4\big) \nonumber \\ &&-(q_1^2-q_3^2)^2(2+q_1^2+q_3^2)\Big]\bigg\} +L(q_1)\bigg\{7+{q_2^2-q_3^2\over 2q_1^2}+{2\over D}\Big( q_2^4+2q_3^4-3q_2^2q_3^2-q_1^2(3q_2^2+2q_3^2)\Big)\nonumber\\ &&+{1\over 2\Sigma}\bigg[{4+q_2^2+q_3^2\over 4+q_1^2}\Big(8(6+q_2^2+q_3^2)-(q_2^2-q_3^2)^2\Big)+q_1^2(12-9q_2^2-5q_3^2-2q_2^4-2q_2^2 q_3^2) \nonumber\\ && -48 -20(q_2^2+q_3^2)+2q_2^4+6q_3^4-16q_2^2q_3^2-q_2^4q_3^2+ q_2^2q_3^4\bigg]\bigg\} +L(q_3)\bigg\{7+{q_2^2-q_1^2\over 2q_3^2}\nonumber\\ &&+{2\over D}\Big( q_2^4+2q_1^4-3q_1^2q_2^2-q_3^2(2q_1^2+3q_2^2)\Big)+{1\over 2\Sigma}\bigg[{4+q_1^2+q_2^2\over 4+q_3^2}\Big(8(6+q_1^2+q_2^2)-(q_1^2-q_2^2)^2\Big)\nonumber\\ && +q_3^2(12-9q_2^2-5q_1^2-2q_2^4-2q_1^2 q_2^2) -48 -20(q_1^2+q_2^2)+2q_2^4+6q_1^4-16q_1^2q_2^2-q_1^2q_2^4+ q_1^4q_2^2\bigg]\bigg\} 
\nonumber\\ && +18 \ln{m_\pi \over \lambda}-{3\over 2}\,,\end{eqnarray}

\begin{eqnarray}\widetilde{\cal  K}_1({\bf q})&\!\!\!=\!\!\!& {3q_2^2\over 2D}(q_2^2-q_1^2-q_3^2) +{q_2^2-q_3^2\over q_1^2}+{q_2^2-q_1^2\over q_3^2}+2J({\bf q})\bigg\{3q_1^2+3 q_3^2-1+{9q_1^2q_2^2q_3^2 \over 4D^2} \Big(q_2^4+2q_2^2(q_1^2+q_3^2)\nonumber \\ &&-3(q_1^2-q_3^2 )^2\Big) + {1\over D}\Big[q_2^2\Big(3q_1^2+3q_3^2 +4q_1^4+4q_3^4+ {75\over 4}q_1^2q_3^2\Big)-(q_1^2-q_3^2)^2(3+4q_1^2+4q_3^2)\Big]\bigg\}\nonumber\\ && +L(q_2)\bigg\{2q_1^2\!+\!2q_3^2\!-\!1 +{9\over 4D^2}\Big[ q_2^4(q_1^4\!+\!q_3^4\!-\!10q_1^2 q_3^2) -q_2^6(q_1^2\!+\!q_3^2) +q_2^2(q_1^2\!+\!q_3^2)(q_1^2\!\!-q_3^2)^2\!-\!(q_1^2\!-\!q_3^2)^4\Big] \nonumber\\ &&+{1\over 4D}\Big[ 37(q_1^2-q_3^2)^2-q_2^2(96+95q_1^2+95q_3^2)\Big] +{2\over \Sigma}
\Big[(q_1^2-q_3^2)^2(4+q_1^2+q_3^2)  -q_2^2\Big(12(1+q_1^2+q_3^2)\nonumber \\ && +2q_1^4+2q_3^4+9 q_1^2q_3^2+q_1^4q_3^2+q_1^2q_3^4 \Big)\Big] \bigg\} +L(q_1)\bigg\{ {q_3^2-q_2^2\over q_1^2}-3 +{9q_2^2 \over 4D^2}\Big[q_1^4 (5q_3^2-q_2^2)+(q_2^2-q_3^2)^3 \nonumber \\ &&+q_1^2(6q_2^2q_3^2-q_2^4-5q_3^4)+ q_1^6\Big]
 +{1\over 4D}\Big[q_1^2(48+71q_2^2+36q_3^2)+48(q_2^2-q_3^2) -13 q_2^4-36q_3^4+49 q_2^2 q_3^2 \Big] \nonumber\\ &&+{1\over \Sigma}\bigg[{4+q_2^2+q_3^2\over 4+q_1^2} (q_2^2-q_3^2)^2+q_1^2(12+13q_2^2+5q_3^2+2q_2^4+2q_2^2 q_3^2)+(6+q_2^2)(2+q_3^2)(q_2^2-q_3^2)\bigg]\bigg\} \nonumber\\ && +L(q_3)\bigg\{ {q_1^2-q_2^2\over q_3^2}-3 +{9q_2^2 \over 4D^2}\Big[q_3^4 (5q_1^2-q_2^2)+(q_2^2-q_1^2)^3 +q_3^2(6q_1^2q_2^2-q_2^4-5q_1^4)+ q_3^6\Big]\nonumber\\ && +{1\over 4D} \Big[q_3^2(48+71q_2^2+36q_1^2) +48(q_2^2-q_1^2) -13 q_2^4-36q_1^4+49 q_1^2 q_2^2 \Big]+{1\over \Sigma}\bigg[ {4+q_1^2+q_2^2\over 4+q_3^2} (q_1^2-q_2^2)^2\nonumber\\ && +q_3^2(12+13q_2^2 +5q_1^2+2q_2^4+2q_1^2 q_2^2)+(6+q_2^2)(2+q_1^2)(q_2^2-q_1^2)\bigg]\bigg\}-15 \ln{m_\pi \over \lambda}+{1\over 4}\,.\end{eqnarray}
 The remaining kernel-functions ${\cal  K}_{2,3}({\bf q})$ and $\widetilde{\cal  K}_{2,3}({\bf q})$ turn out be extremely lengthy, mainly because in their expansions with respect to $J({\bf q})$ and $L(q_j)$ the coefficients involve yet higher powers of $1/D$.  Nevertheless we exhibit some essential parts of their compositions, which read:
\begin{eqnarray} {\cal K}_2({\bf q}) &\!\!\!=\!\!\!&\text{comb}\big\{J({\bf q}),L(q_2),L(q_1),L(q_3)\big\}+{15(q_1 q_2 q_3)^2 \over 32 D^2}\Big[3(q_1^2-q_3^2)^2-q_2^4-2q_2^2(q_1^2+q_3^2)\Big] \nonumber\\ &&  +{1\over 32 D}\Big[2(q_1^2-q_3^2)^2\big(13q_1^2+13q_3^2-6\big)+q_2^2 \big(12q_1^2+12q_3^2-26q_1^4-26q_3^4 -133q_1^2q_3^2\big)\Big] \nonumber\\ && +{q_2^2-q_3^2\over 3q_1^2}+ {q_2^2-q_1^2\over 3q_3^2}-{1\over 4}\Big(30+3q_1^2+3q_3^2+{17\over 6}q_2^2\Big) \ln{m_\pi \over \lambda} +{55\over 48} -{83\over 160}(q_1^2+q_3^2)+{433\over 2880} q_2^2\,,  \nonumber\\ &&\end{eqnarray} 
\begin{eqnarray} \widetilde{\cal K}_2({\bf q}) &\!\!\!=\!\!\!&\text{comb}\big\{J({\bf q}),L(q_2),L(q_1),L(q_3)\big\}+{1\over D}\Big[ q_2^2(q_1^4\!+\!q_3^4\!+\!3q_1^2 q_3^2) -(q_1^2\!-\!q_3^2)^2(q_1^2\!+\!q_3^2)\Big]  +{2\over 3q_1^2}(q_3^2-q_2^2)\nonumber\\ &&  + {2\over 3q_3^2}(q_1^2-q_2^2)+\Big[36+{1\over 3}(13q_1^2+13q_3^2+11 q_2^2) \Big] \ln{m_\pi \over \lambda} -{16\over 3} -{1\over 72}\Big[55(q_1^2+q_3^2)+161 q_2^2\Big]\,, \end{eqnarray} 
\begin{eqnarray} {\cal K}_3({\bf q}) &\!\!\!=\!\!\!&\text{comb}\big\{J({\bf q}),L(q_2),L(q_1),L(q_3)\big\}+{1\over 2D}\Big[ q_2^2(q_1^4+q_3^4+3q_1^2 q_3^2) -(q_1^2-q_3^2)^2(q_1^2+q_3^2)\Big] \nonumber\\ &&  +{2\over 3q_1^2}(q_2^2-q_3^2) + {2\over 3q_3^2}(q_2^2-q_1^2)-\Big[90+{17\over 2}(q_1^2\!+\!q_3^2)+{61\over 6} q_2^2\Big] \ln{m_\pi \over \lambda} +{11\over 3}+{67\over 48}(q_1^2\!+\!q_3^2)+ {775\over 144}q_2^2\,, \nonumber\\ &&\end{eqnarray} 
\begin{eqnarray} \widetilde{\cal K}_3({\bf q}) &\!\!\!=\!\!\!&\text{comb}\big\{J({\bf q}),L(q_2),L(q_1),L(q_3)\big\}+{15(q_1 q_2 q_3)^2 \over 16 D^2}\Big[3(q_1^2-q_3^2)^2-q_2^4-2q_2^2(q_1^2+q_3^2)\Big] \nonumber\\ &&  +{3\over 16 D}\Big[2(q_1^2-q_3^2)^2\big(7q_1^2+7q_3^2+6\big)-q_2^2 \big(12q_1^2+12q_3^2 +14 q_1^4+14q_3^4 +63q_1^2q_3^2\big)\Big]  +{4\over q_1^2}(q_3^2-q_2^2)
\nonumber\\ && + {4\over q_3^2}(q_1^2-q_2^2)+\Big(105+9q_1^2+9q_3^2+{43\over 4}q_2^2\Big) \ln{m_\pi \over \lambda}-{39\over 8}-{57\over 20}(q_1^2+q_3^2)-{3359\over 480}q_2^2\,, \end{eqnarray} 
where $\text{comb}\big\{J({\bf q}),L(q_2),L(q_1),L(q_3)\big\}$ stands for a linear combination with expansion coefficients, that are rational functions of  $q_1, q_2, q_3$. The constant and quadratic polynomial at the end of each formula  are specific for our regularization method with a euclidean cutoff $\lambda$. The knowledge of the constant coefficients of $\ln(m_\pi/\lambda)$ allows one to verify that no contributions to $\bar E_n(\rho_n)$ of the form $g_A^4c_{1,2,3,4}m_\pi^2 k_n^6/(2\pi f_\pi)^6 \ln(m_\pi/\lambda)$  exist. Full expressions for the kernel-functions ${\cal K}_{2,3}({\bf q})$ and $\widetilde{\cal K}_{2,3}({\bf q})$ can be obtained from the author upon request.
\section{Subleading three-nucleon contact-potential}
The subleading three-nucleon contact potential (appearing at N$^4$LO) has been reexamined recently in ref.~\cite{3Ncontact}. Its corrected version depends quadratically on momenta and it involves 13 parameters, called $E_1,\dots,E_{13}$.
The full expression for the  subleading 3N contact interaction reads:
\begin{eqnarray} V_\text{3N}&\!\!\!=\!\!\!& -E_1\,\vec q_1^{\,2} -E_2\, \vec q_1^{\,2}\vec\tau_1\!\cdot \!\vec\tau_2-E_3 \,\vec q_1^{\,2}\vec\sigma_1\!\cdot \!\vec\sigma_2 -E_4 \,\vec q_1^{\,2}\vec\sigma_1\!\cdot \!\vec\sigma_2 \,\vec\tau_1\!\cdot \!\vec\tau_2-E_5(3 \vec\sigma_1\!\cdot \!\vec q_1\, \vec\sigma_2\!\cdot \!\vec q_1-\vec q_1^{\,2}\vec\sigma_1\!\cdot \!\vec\sigma_2 )\nonumber\\ && -E_6(3 \vec\sigma_1\!\cdot \!\vec q_1\, \vec\sigma_2\!\cdot \!\vec q_1-\vec q_1^{\,2}\vec\sigma_1\!\cdot \!\vec\sigma_2 )\vec\tau_1\!\cdot \!\vec\tau_2 +{i\over 4}E_7(\vec\sigma_1\!+ \!\vec\sigma_2 )\!\cdot \!\vec q_1\!\times \!(\vec p_1\!+\!\vec p_1\,\!\!'\!-\!\vec p_2\!-\!\vec p_2\,\!\!')\nonumber\\ &&  +{i\over 4}E_8(\vec\sigma_1\!+ \!\vec\sigma_2 )\!\cdot \!\vec q_1\!\times \!(\vec p_1\!+\!\vec p_1\,\!\!'\!-\!\vec p_2\!-\!\vec p_2\,\!\!')\vec\tau_2\!\cdot \!\vec\tau_3 -E_9 \,\vec\sigma_1\!\cdot \!\vec q_1\, \vec\sigma_2\!\cdot \!\vec q_2 -E_{10}\, \vec\sigma_1\!\cdot \!\vec q_1\, \vec\sigma_2\!\cdot \!\vec q_2\, \vec\tau_1\!\cdot \!\vec\tau_2 \nonumber\\ &&-E_{11}\, \vec\sigma_2\!\cdot \!\vec q_1\, \vec\sigma_1\!\cdot \!\vec q_2 -E_{12}\, \vec\sigma_2\!\cdot \!\vec q_1\, \vec\sigma_1\!\cdot \!\vec q_2\, \vec\tau_1\!\cdot \!\vec\tau_2 -E_{13} \,\vec\sigma_2\!\cdot \!\vec q_1\, \vec\sigma_1\!\cdot \!\vec q_2\, \vec\tau_1\!\cdot \!\vec\tau_3 \,.  \end{eqnarray} 
The evaluation of the closed 3-ring, 2-ring, and 1-ring diagrams in Fig.\,2 with this  $V_\text{3N}$ gives contributions to the energies per particle that are proportional to the eighth power of the respective Fermi-momentum:
\begin{eqnarray} \bar E(\rho) &\!\!\!=\!\!\!& {k_f^8\over 10\pi^4}\big(2E_1+2E_2+2E_3+6E_4-E_9-3E_{10} -E_{11} -3E_{12}+E_{13}\big)\,,  \\  \bar E_n(\rho_n) &\!\!\!=\!\!\!& {k_n^8\over 30\pi^4}\big(2E_1+2E_2-E_9-E_{10} -E_{11} -E_{12}-E_{13}\big)\,,  \end{eqnarray}
where the tensor terms $\sim E_{5,6}$ and spin-orbit terms $\sim E_{7,8}$ have obviously dropped out at first order.
\section*{Appendix: Leading order chiral three-nucleon force}
For the sake of completeness we reproduce here also the results for $\bar E(\rho)$ and $\bar E_n(\rho_n)$ as obtained from leading order chiral 3N-interaction at N$^2$LO. The two-pion exchange component $\sim c_{1,3,4}$ gives:
\begin{eqnarray}\bar E(\rho)^{2r} &\!\!\!=\!\!\!& {g_A^2 m_\pi^6 \over (2\pi f_\pi)^4} \bigg\{(12c_1-10c_3) u^3 \arctan 2u -{4c_3 \over 3}u^6+6(c_3-c_1)u^4+(3c_1-2c_3)u^2 \nonumber\\ &&+\Big[ {1\over 4}(2c_3-3c_1)+{3u^2\over 2}(3c_3-4c_1)\Big] \ln(1+4u^2)\bigg\}\,,\end{eqnarray}
\begin{eqnarray}\bar E_n(\rho_n)^{2r} &\!\!\!=\!\!\!& {g_A^2 m_\pi^6 \over (2\pi f_\pi)^4} \bigg\{ \Big(2c_1-{5c_3\over 3}\Big) u^3 \arctan 2u -{2c_3 \over 9}u^6+(c_3-c_1)u^4+\Big({c_1\over 2}-{c_3\over 3}\Big)u^2 \nonumber\\ &&+\Big[ {1\over 24}(2c_3-3c_1)+{u^2\over 4}(3c_3-4c_1)\Big] \ln(1+4u^2)\bigg\}\,,\end{eqnarray}
\begin{equation}\bar E(\rho)^{1r} = {3g_A^2 m_\pi^6 \over (4\pi f_\pi)^4u^3}\int_0^u\!\!dx\Big\{ 3c_1[G(x)]^2+\Big({c_3\over 2}-c_4\Big) [G_s(x)]^2+(c_3+c_4)[G_t(x)]^2 \Big\}\,, \end{equation}
\begin{equation}\bar E_n(\rho_n)^{1r} = {g_A^2 m_\pi^6 \over (4\pi f_\pi)^4u^3}\int_0^u\!\!dx\Big\{ 3c_1[G(x)]^2+{c_3\over 2} [G_s(x)]^2+c_3[G_t(x)]^2 \Big\}\,. \end{equation}
On the other hand $1\pi$-exchange combined with the 4N$1\pi$-contact coupling produces the result:
\begin{equation}\bar E(\rho) = {g_Ac_D m_\pi^6 \over (2\pi f_\pi)^4\Lambda_\chi} \bigg\{ 
{u^6\over 3}-{3u^4\over 4}+{u^2\over 8} +u^3 \arctan 2u -{1\over 32}(1+12u^2)\ln(1+4u^2) \bigg\}\,, \end{equation}
and the six-nucleon contact term leads to the $\rho^2$-piece:
\begin{equation}\bar E(\rho) =-{c_E \,k_f^6 \over 12\pi^4 f_\pi^4\Lambda_\chi} \,, \end{equation}
with no further contribution to pure neutron matter.
\vspace{-0.3cm}\subsection*{Acknowledgement}
I thank H. Krebs for providing me files with the non-polynomial parts of the 3N-ring interaction. 
\end{document}